\documentstyle[aps,preprint,epsfig]{revtex}
\begin{document}
\draft

\title{
Isoscalar-Isovector Interferences
in $\bbox{\pi N\to N e^+e^-} $ Reactions as a Probe
of Baryon Resonance Dynamics}

\author{
{\sc A.I.~Titov}$^{a,b}$,
{\sc B.~K\"ampfer}$^{a}$
}

\address{
$^a$ Forschungszentrum Rossendorf, PF 510119, 01314 Dresden,
Germany\\
$^b$ Bogolyubov Laboratory of Theoretical Physics, JINR,
Dubna 141980, Russia
}

\maketitle

\begin{abstract}
The isoscalar-isovector ($\rho-\omega$) interferences in
the exclusive reactions
$\pi^-p\to n e^+e^-$ and $\pi^+n \to p e^+e^-$
near the $\omega$ threshold
leads to a distinct difference of the dielectron
invariant mass distributions depending on
beam energy. The strength of this effect is determined
by the coupling of resonances to the nucleon vector-meson channels and other
resonance properties. Therefore, a combined analysis of these
reactions can be used as a tool for determining 
the baryon resonance dynamics.\\[3mm]
{\it keywords:} hadron reactions; baryon resonances;
rho and omega interference.

\end{abstract}

\section{Introduction} 

The study of
dielectron production in hadron and heavy-ion reactions
addresses various issues of general interest.
In heavy-ion collisions the dileptons are
considered as a tool for accessing in-medium
modifications of vector mesons.
For example, at relativistic energies, the behavior of the
$\rho$ meson attracted much attention because the dilepton data
\cite{CERES} point to a reshuffling of strength
in a hot, meson-dominated medium \cite{rho_shift,Rapp_Wambach}.
This has been discussed in the wider context of chiral symmetry restoration
(cf.\ \cite{Rapp_Wambach,Brown_Rho_review}),
QCD sum rules (cf.\ \cite{Shifman_et_al}),
and hadronic models (cf.\ \cite{Weise}).
Likewise, the dielectron production in heavy-ion collisions
at beam energies of a $\sim 1$ AGeV is interesting due to similar reasons.
Also here the dielectron channel is considered as an appropriate tool
for studying in-medium modification of vector mesons in
a baryon-dense medium.
After the first round of experiments with DLS \cite{DLS},
the HADES spectrometer at the heavy-ion
synchrotron SIS at GSI/Darmstadt \cite{HADES},
beginning now with experiments, is built
to verify these predictions related to fundamental symmetry properties
of strong interaction physics. The experimental feasibilities at HADES
(e.g., the disposal of beams of pions, protons and a wide range of nuclei)
triggered an enhanced activity in this field.

Clearly, for an understanding of dielectron spectra in hadron-nucleus
and heavy-ion collisions the elementary hadronic reaction channels
must be under control. Here 
the reactions $NN \to X e^+ e^-$ and $\pi N \to X e^+ e^-$
occupy a highly important place in the dielectron physics.
These reactions serve as a necessary input
for kinetic approaches
(cf.\ \cite{rho_shift,EBCM99}).
But on the other hand, they are interesting for themselves because 
they are  mainly related to the baryon resonance dynamics.
Different facets of the manifestation of baryon
resonances in $\pi N$ collisions have been analyzed in
Refs.~\cite{EBCM99,FP97,F98,LWF99,SLF00,M_P_L,TKR01,Lykasov}.
Particularly interesting
are such recent theoretical approaches as in 
\cite{LWF99,M_P_L}
which attempt a unifying description of meson-nucleon interactions.

The quantum interference in $e^+e^-$ decays of 
intermediate $\rho$ and $\omega$ mesons produced
in the exclusive reaction $\pi^- p \to n \rho (\omega) \to n e^+ e^-$ 
has been first discussed in Ref.~\cite{SLF00},
and the first round of HADES experiments will experimentally address this
problem \cite{HADES}.
Here, we would like to emphasize
that the $\rho-\omega$ interference in dielectron production has also
an another interesting aspect:
The interference may be used as a tool for studying the isoscalar part
of the electromagnetic current in the resonance region, what is
rather difficult to do by another methods. Varying the dilepton
invariant mass $M$ one can test low-lying baryon resonances which are
deeply subthreshold for on-shell omega production. Of course,
the contribution of the isoscalar part 
(i.e., the virtual $\omega$ production)
is much smaller than the dominant isovector 
(i.e., virtual $\rho$ production)
at $M \neq m_\omega$ but it may be clearly seen in the $\rho-\omega$
interference which is proportional to the difference of the
$ e^+ e^-$ cross sections in $\pi^-p$ and $\pi^+n$ collisions.

Indeed, since the electromagnetic
current is the sum of isoscalar and isovector components
\cite{Kroll_Lee_Zumino}, 
the invariant
amplitude of the reaction $\pi^-p\to n e^+e^-$ with total isospin
$I = \frac12$ may be expressed as
\begin{eqnarray}
T^{\pi^-p\to n e^+e^-} \propto T^{\rm scalar} + T^{\rm vector},
\end{eqnarray}
where, according to the vector dominance model, the isoscalar (isovector)
contributions may be identified with $\omega$ ($\rho$) meson
intermediate states, i.e.
$T^{\rm scalar} \propto T^\omega$ and $T^{\rm vector} = T^\rho$.
A rotation by 180$^o$ around the $y$-axis in isospin space leads
to the transformations 
$\vert p \rangle \to - \vert n \rangle$, 
$\vert n \rangle \to \vert p \rangle$, 
$\vert \pi^- \rangle \to - \vert \pi^+ \rangle$,
$ \vert \omega \rangle \to \vert \omega \rangle$, 
$\vert \rho^0 \rangle \to - \vert \rho^0 \rangle$ 
and, therefore, one gets
\begin{eqnarray}
T^{\pi^+n \to p e^+ e^-} \propto T^{\rm scalar} - T^{\rm vector}.
\end{eqnarray}
That means, the quantum interferences in the reactions 
$\pi^- p$ and $\pi^+ n$ are
different, and these differences might be well observable
in the vicinity of the $\omega$ resonance peak.

In Ref.~\cite{TKR01} the $\omega$ production in $\pi N$ interactions 
has been analyzed within an approach based on tree level diagrams
and effective Lagrangians.
A strong contribution in the near-threshold energy region is found 
to stem from the $s$ and $u$ channels of nucleon
and baryon resonances amplitudes. In the present work we will consider these
dominant amplitudes as depicted in Fig.~1.
(The restriction to $s$ and $u$ channels,
and the exclusion of the $t$ channel, 
is in line with the concept of duality.)
We will account for the resonances
with mass $M_{B^*} \le 1.72$ GeV
($B^* = N, N^*, \Delta$).
This means that, together with the
$T^\omega \pm T^\rho$ interferences one has to consider the strong
"internal" interferences within the $\omega$ and $\rho$ channels separately
which are in turn different in both channels.
Therefore, the proper choice of the $\pi NB^*$, $\omega NN^*$,
and $\rho NB^*$ coupling constants and their phases
becomes the central problem. To demonstrate the  
$T^\omega \pm T^\rho$ interferences within a concise framework
we rely mainly on \cite{RB00},
where the relevant coupling constants are expressed in terms of
the corresponding couplings to the nucleon by using a quark model.
We will also briefly discuss the possibility to use the known partial widths
of $B^* \to N \rho$ decays to fix the absolute values
of $\rho N B^*$ couplings.
Our approach highlights the role of the coupling of subthreshold
resonances to the $N \rho$ and $N \omega$ systems
(cf.\ \cite{LWF99,M_P_L,Manley} for discussion and further references).

Our paper is organized as follows.
In Section II, we define the effective
Lagrangians, derive expressions for invariant
amplitudes of the processes shown in Fig.~1
and discuss the parameter fixing.
In Section III the results of numerical
calculations and predictions are presented.
The summary is given in Section IV.
In the Appendices we show explicit expressions
of effective Lagrangians and invariant amplitudes.

\section{Amplitudes} 

The differential cross section of the reaction
$\pi N \to N e^+ e^-$ averaged over the
azimuthal angle of the electron is defined as
\begin{eqnarray}
\frac{d\sigma}{d\,\Omega d\Omega_e dM^2}
=\frac{\alpha M^2}{8\pi^2}\,
\left[\Sigma_{\parallel}\sin^2\Theta +
\Sigma_{\perp}(1+\cos^2\Theta)\right],
\label{CS_ee}
\end{eqnarray}
where $\Omega_e$ and $\Theta$ are the solid and polar angles
of the electron, $\Omega$ and $\theta$ denote the solid and polar angles
of the dielectron {in the center of mass system of the entrance channel}, and
$M$ stands for the invariant $e^+ e^-$ mass.
The longitudinal and transversal distributions
$\Sigma_{\parallel, \perp}$ read
\begin{eqnarray}
\Sigma_{\parallel}
& = &
\frac{1}{128\pi^2s} \, \frac{|{\bf q}|}{|{\bf k}|}
\sum_{s_i,s_f}
\left\vert
\frac{f_\rho {T^{\lambda=0}_\rho}_{s_i,s_f}}{M^2-m_\rho^2+im_\rho\Gamma_\rho}
+\frac{f_\omega {T^{\lambda=0}_\omega}_{s_i,s_f}}
{M^2-m_\omega^2+im_\omega\Gamma_\omega}
\right\vert^2 , \nonumber\\
\Sigma_{\perp}
& = &
\frac{1}{128\pi^2s}\,\frac{|{\bf q}|}{|{\bf k}|}
\sum_{s_i,s_f}
\left\vert
\frac{f_\rho {T^{\lambda=1}_\rho}_{s_i,s_f}}{M^2-m_\rho^2+im_\rho\Gamma_\rho}
+\frac{f_\omega {T^{\lambda=1}_\omega}_{s_i,s_f}}
{M^2-m_\omega^2+im_\omega\Gamma_\omega}
\right\vert^2 ,
\label{S_ee}
\end{eqnarray}
where $k = (E_\pi,\, {\bf k})$ and $q=(E_V, {\bf q})$ are the
four-momenta of the pion and the dielectron (or the intermediate vector
meson) in the center of mass system. We denote
the four-momenta of the initial (target) and final (recoil) nucleons by
$p$ and $p'$; $s = (p+k)^2$ is the usual Mandelstam variable.
${T^\lambda_{\rho(\omega)}}_{s_i,s_f}$ stands for the invariant
amplitude of the virtual vector meson $\rho$ ($\omega$) production with
polarization $\lambda$ and nucleon spin projections $s_i, s_f$; $m_V$
(with $V = \omega, \rho^0$) is the vector meson mass, $f_V$ denote the
coupling constants of the $V\to e^+e^-$ decays, and $\Gamma_V$ are the
total decay widths. For the $\omega$ meson,
$\Gamma_\omega = 8.41$ MeV \cite{PDG98}, while
for the wide $\rho$ meson we use the energy dependent width
\begin{eqnarray}
{\Gamma_\rho} = {\Gamma_\rho}^0
\left[\frac{M^2-4m_\pi^2}{m_\rho^2-4m_\pi^2},
\right]^{\frac32},
\end{eqnarray}
keeping the strongest $M$ dependence which comes from the corresponding
$\rho \pi\pi$ Lagrangian, 
with ${\Gamma_\rho}^0 = 150.7$ MeV \cite{PDG98}.

The differential invariant mass distribution integrated over
$d\Omega_e$ reads
\begin{eqnarray}
\frac{d\sigma}{d\,\Omega dM^2}
=\frac{\alpha M^2}{3\pi}\,
\left[\Sigma_{\parallel} +
2\Sigma_{\perp}\right].
\label{CSi_ee}
\end{eqnarray}

\subsection{Effective Lagrangians} 
\label{effective_L}

Calculating the invariant amplitudes for the basic processes shown
in Fig.~1
we use the following effective interaction Lagrangians
in symbolic notation
\begin{eqnarray}
{\cal L}_{\pi NB^*}
=
f_{\pi NN} \bar\psi_N\,{\cal F}_{N}\bbox{\pi\cdot t}\psi_N
&+&\sum_i f_{\pi NB^*_i}\bar\psi_N\, {\cal F}_{i}\bbox{\pi\cdot t}\psi^i
+\sum_i f_{\pi NB^*_i}\bar\psi_N\, {\cal F}_{i}^\alpha\bbox{\pi\cdot t}
\psi^i_{\alpha}\nonumber\\
&+&\sum_i f_{\pi NB^*_i}\bar\psi_N\, {\cal F}_{i}^{\alpha\beta}
\bbox{\pi \cdot t}
\psi^i_{\alpha\beta}  + {\rm h.c.},
\label{LpiB}
\end{eqnarray}
\begin{eqnarray}
{\cal L}_{\omega NN^*}
=
g_{\omega NN} \bar\psi_N\,{\cal G}^\mu_N\psi_N\omega_\mu
&+&\sum_i g_{\omega NN^*_i}\bar\psi_N\, {\cal G}^\mu_{i}\psi^i\omega_\mu
+\sum_i g_{\pi NN^*_i}\bar\psi_N\, {\cal G}_{i}^{\mu\alpha}
\psi^i_{\alpha}\omega_\mu  \nonumber\\
&+&\sum_i g_{\pi NN^*_i}\bar\psi_N\, {\cal G}_{i}^{\mu\alpha\beta}
\psi^i_{\alpha\beta}\omega_\mu    + {\rm h.c.},
\label{LomegaB}
\end{eqnarray}
\begin{eqnarray}
{\cal L}_{\rho NB^*}=
g_{\rho NN} \bar\psi_N\,{\cal G}^\mu_{N}
\bbox{\rho}_\mu \bbox{\cdot t}\psi_N
&+&\sum_i g_{\rho NB^*_i}\bar\psi_N \, {\cal G}^\mu_{i}
\bbox{\rho}_\mu \bbox{\cdot t}\psi^i
+\sum_i g_{\rho NB^*_i}\bar\psi_N \, {\cal G}_{i}^{\mu\alpha}
\bbox{\rho}_\mu \bbox{\cdot t}
\psi^i_{\alpha}\nonumber\\
&+&\sum_i g_{\rho NB^*_i}\bar\psi_N\,
{\cal G}_{i}^{\mu\alpha\beta}
\bbox{\rho}_\mu \bbox{\cdot t}
\psi^i_{\alpha\beta}  + {\rm h.c.},
\label{LrhoB}
\end{eqnarray}
where ${\bbox \pi}$, $\bbox{\rho}_\mu$ and $\omega_\mu$, are the pion,
rho and omega meson fields,
$\psi_N$, $\psi^i$, $\psi^i_{\alpha}$ and $\psi^i_{\alpha\beta}$
stand for the nucleon, spin-$\frac12$, spin-$\frac32$ and spin-$\frac52$
baryon resonances, respectively. For spin-$\frac32$ and $\frac52$
fields we use Rarita-Schwinger field operators.
$\alpha, \beta, \gamma, \cdots \mu, \nu,\cdots $ are Lorentz indices;
$i$ enumerates the corresponding baryon states.
The isospin operator $\bbox{t}$ is just Pauli's matrix $\bbox{\tau}$
for the nucleon and nucleon resonances with isospin $\frac12$, while
for isospin-$\frac32$ it is the transition matrix $\bbox{\chi}$ for
delta-resonances, see \cite{RB00}.
In the isoscalar amplitude we include the contribution of
the nucleon ($N$) and the 8 resonances ($N^*$)
$P_{11}(1440)$, $D_{13}(1520)$, $S_{11}(1535)$, $S_{11}(1650)$,
$D_{15}(1675)$, $F_{15}(1680)$, $D_{13}(1700)$, $P_{13}(1720)$
(the neglect of $P_{11}(1710)$ is motivated in \cite{TKR01}).
For the isovector amplitude we consider
{these states and additionally} the 4 $\Delta$ states up to 1700 MeV
(all together $B^*$)
$P_{33}(1232)$, $P_{33}(1600)$, $S_{31}(1620)$, $D_{33}(1700)$.
The explicit form of the employed effective Lagrangians is listed
in Appendix A, where the symbols ${\cal G}_{\cdots}^{\cdots}$
and ${\cal F}_{\cdots}^{\cdots}$ are resolved.

\subsection{Invariant amplitudes} 

The isoscalar invariant amplitude is the coherent sum
of nucleon and resonance  channels in the following form
(the nucleon spin projections are now suppressed)
\begin{eqnarray}
T_\omega^{ \, \lambda}(N)
& = &
g_{\omega NN}\,\frac{f_{\pi NN}}{m_\pi}\,
\bar{u}(p') \, \left[{\cal A}^{(\omega) \mu}_s (N) +
{\cal A}^{(\omega) \mu}_u (N) \right]\,
u(p) \, \varepsilon^{* \, \lambda}_\mu \, I_\omega, \nonumber\\
T_\omega^{ \, \lambda}(N^*)
& = &
g_{\omega NN^*}\,\frac{f_{\pi NN^*}}{m_\pi}\,
\bar{u}(p') \, \left[{\cal A}^{(\omega) \mu}_s({N^*}) +
{\cal A}^{(\omega) \mu}_u (N^*)\right]\,
u(p) \, \varepsilon^{* \, \lambda}_\mu \, I_\omega,
\label{T_om_N*}
\end{eqnarray}
where $\varepsilon^\lambda_\mu$ is
the polarization four-vector for a spin-1 particle with spin projection
$\lambda$, four-momentum $p=(E,{\bf p})$ and mass $m$
\begin{eqnarray}
\varepsilon^\lambda(p) = \left( \,
\frac{{\bbox \epsilon}^\lambda\cdot{\bf p}}{m},\,\,
{\bbox\epsilon}^\lambda +
\frac{{\bf p}\,({\bbox \epsilon}^\lambda\cdot{\bf p})}{m ( E + m)}\,
\right),
\end{eqnarray}
with the three-dimensional polarization vector ${\bbox \epsilon}$
with components
${\bbox \epsilon}^{\pm 1} = \mp \frac{1}{\sqrt2} (\,1,\,\,\pm i,\,\,0\,)$,
${\bbox \epsilon}^{0} = (\,0,\,\,0,\,\,1\,)$.

The isovector invariant amplitude has a slightly
different form because of the corresponding isospin factors,
\begin{eqnarray}
T_\rho^\lambda (N)
& = &
g_{\rho NN}\,\frac{f_{\pi NN}}{m_\pi}\,
\bar{u}(p') \, \left[{\cal A}^{(\rho) \mu}_s (N) -
{\cal A}^{(\rho) \mu}_u (N) \right]\,
u(p) \, \varepsilon^{* \, \lambda}_\mu \, I_\rho(N), \nonumber\\
T_\rho^\lambda (N^*)
& = &
g_{\rho NN^*}\,\frac{f_{\pi NN^*}}{m_\pi}\,
\bar{u}(p') \, \left[{\cal A}^{(\rho) ^\mu}_s (N^*) -
{\cal A}^{(\rho)^\mu}_u (N^*) \right]\,
u(p) \, \varepsilon^{* \, \lambda}_\mu \, I_\rho(N),\nonumber\\
T_\rho^\lambda (\Delta^*)
& = &
g_{\rho N\Delta}\,\frac{f_{\pi N\Delta}}{m_\pi}\,
\bar{u}(p') \, \left[{\cal A}^{(\rho)^\mu}_s (\Delta) +
{\cal A}^{(\rho) \mu}_u (\Delta) \right]\,
u(p) \, \varepsilon^{* \, \lambda}_\mu \, I_\rho(\Delta),
\label{T_rho_B*}
\end{eqnarray}
where the isospin factor $I_\rho(N) = -\sqrt{2}$ for
the reaction $\pi^-p \to n e^+ e^-$
($+ \sqrt{2}$  for $\pi^+n \to pe^+ e^-$) and
$I_\rho(\Delta)=\sqrt{2}/3$.
The $s$ and $u$ channel operators 
${\cal A}^{(\rho, \omega) \mu}_s$ and 
${\cal A}^{(\rho, \omega) \mu}_u$
in Eq.~(\ref{T_om_N*})
are defined by the effective Lagrangians
of Eqs.~(\ref{LpiB}, \ref{LomegaB}) and listed in Appendix B.
$I_\omega=\sqrt{2}$ is the isospin factor.

Following the previous studies \cite{TKR01,FM98,TKR00}
we assume that the vertices must be dressed
by form factors for off-shell baryons
\begin{eqnarray}
F_{B^*}(r^2) =
\frac{\Lambda_B^{* 4}}{\Lambda_B^{* 4} + (r^2 - M^2_{B^*})^2 },
\label{cutN}
\end{eqnarray}
where $r$ is the four-momentum of the virtual baryons $B^*$
with mass $M_{B^*}$.
Eq.~(\ref{cutN}) represents the simplest form being
symmetric in the $s$ and $u$ channels. The form factor
is positive and decreases with increasing off-shellness in both channels.

An analysis of Eqs.~(\ref{T_om_N*} - \ref{T_rho_B*})
shows that
(i) the interference between $s$ and $u$ channels is different for the
$\omega$ and $\rho$ production amplitudes,
(ii) an additional difference comes from
the different values and phases of the couplings $g_{VNN^*}$
for the same resonances, and
(iii) the $\rho-\omega$ interference is different for
$\pi^-p$ and $\pi^+ n$ interactions,
as already anticipated in Eqs.~(1, 2).

\subsection{Fixing parameters} 

The coupling constants $f_V$ of the decays
$V = \rho , \omega \to e^+ e^-$ in Eq.~(\ref{S_ee})
are related to the corresponding decay widths as
\begin{eqnarray}
f_V^2=\frac{3\Gamma_{V\to e^+e^-}}{\alpha \, m_V}.
\end{eqnarray}
Using $\Gamma_{\rho\to e^+e^-}=6.77$ keV and
$\Gamma_{\omega\to e^+e^-}=0.60$ keV
\cite{PDG98} one gets $f_\rho=0.06$ and $f_\omega=0.0177$.
The nucleon and nucleon resonance amplitudes in Fig.~1
are determined by the couplings
$f_{\pi NN}$,
$f_{\pi NB^*}$,
$g_{\omega NN}$,
$g_{\omega NN^*}$,
$g_{\rho NN}$,
$g_{\rho NB^*}$,
$g_{VNN}$ and $\kappa_{VNN}$,
the resonance widths $\Gamma^0_{B^*}$
the branching ratios $B^\pi_{B^*}$,
and the cut-offs $\Lambda_B$.
For the coupling constant $f_{\pi NN}$ we use the standard value
$f_{\pi NN}=1.0$ \cite{RB00,Bonn}.
For the $\omega NN$ coupling we use the values
$g_{\omega NN} = 10.35$ and $\kappa_{\omega NN} = 0$
determined recently in \cite{RB00,RSY99}.
For the $\rho NN$ coupling we use the value
$g_{\rho NN} = 3$ and $\kappa_{\rho NN}=6.1$~\cite{RB00,Bonn}.

The values of coupling constants $f_{\pi NB^*}$ are determined from
a comparison of calculated decay widths $\Gamma_{N^*\to N\pi}$
with the corresponding experimental values \cite{PDG98}.
The corresponding signs are taken in accordance
with the quark model prediction of Ref.~\cite{RB00}.

The values of coupling constants $g_{V NB^*}$ follow from
$g_{V NB^*} = [g_{V NB^*}/g_{V NN}] g_{V NN}$,
where the ratio $[g_{V NB^*}/g_{V NN}]$ is determined by
the quark model calculation of Ref.~\cite{RB00}. 
In subsection \ref{III.B} we contrast this choice of the parameters
$g_{V NB^*}$ with another one.

The yet undetermined 13 cut-off parameters $\Lambda_{B^*}$
in Eq.~(\ref{cutN}) are reduced to one by making the natural assumption
\begin{eqnarray}
\Lambda_N &=& \Lambda_{B^*}\equiv\Lambda_B.
\label{LB}
\end{eqnarray}
The  total cross section of real $\omega$ production in the near
threshold region is reproduced by choosing
$\Lambda_B=0.66$ GeV \cite{TKR01}.

\section{Results} 

\subsection{Using coupling parameters from \protect\cite{RB00}} 

Similar to our previous study of $\omega$ production \cite{TKR01}
we use the coupling strengths and phases from \cite{RB00}.
For convenience we show in Table~1
all the coupling constants, decay widths and branching ratios
used in our calculation.
(The masses, decay widths and branching ratios in Table~1
represent the averages in \cite{PDG98}.)
The results of our full calculation of the
differential cross section as a function of dielectron invariant mass
are shown in Fig.~2 for the reaction $\pi^- p \to n e^+ e^-$
at two energies, $s^{1/2} = $1.6 and 1.8 GeV. Here and later on,
the calculations have been done for the dielectron (or virtual vector meson)
production at $\theta=30^o$
{in the corresponding center of mass system},
except for particular cases which are mentioned explicitly below.
We also show separately the contributions of the $\omega$ and $\rho$
channels. At an energy of $s^{1/2}=1.8$ GeV
(see Fig.~2, right panel),
which is about 80 MeV above the $\omega$
production threshold, one can see the sharp $\omega$ resonance peak
at $M \simeq m_\omega$.
Away of the $\omega$ peak position one observes a strong decrease of the
$\omega$ contribution as compared with
the fairly flat $\rho$ contribution, at the exhibited scale.
In contrast, for an energy sufficiently below the $\omega$
threshold (Fig.~2, left panel), also the $\omega$
contribution is a smooth function of $M$ but below the $\rho$
contribution. This is because the suppression of the resonance factor
in Eq.~(\ref{S_ee}) at $M \neq m_\omega$ is much stronger for $\omega$.

For the reaction $\pi^+ n \to p e^+ e^-$, the shape
of the invariant mass distribution
is similar and, therefore, is not displayed here.
But the absolute values of the corresponding
total distributions are different, and this difference is shown in Fig.~3,
where the reactions $\pi^-p\to ne^+e^-$ and $\pi^+n\to pe^+e^-$
are compared.
The difference reaches a factor up to three
and depends on both the energy and the invariant mass.
At low energy the cross section for the reaction
$\pi^-p$ is smaller, while at higher energy it is greater
than that for $\pi^+n$ interactions.
The reason of this effect is the difference in $\rho - \omega$
interferences in the two reactions and a different role of individual
baryon resonances depending on the initial energy.
In order to get insight into the resonance dynamics, 
in Figs.~4 and 5 we show
the contribution of each resonance separately for $\rho$ (left panels)
and $\omega$ (right panels) channels as a function of the
dielectron  production angle for the reaction $\pi^- p \to n e^+ e^-$.
Most transparent is the situation for the $\omega$ channel.
One can see that dominant contributions come from
$S_{11}(1535)$ and $S_{11}(1650)$ resonances. For $\omega$
production their phases are opposite, while for $\rho$
production they are the same. At low energy
(see Fig.~4)
the contribution of $S_{11}(1535)$ is greater and taking into account the
additional isospin factor $I_{\rho}$ in Eq.~(\ref{T_rho_B*}) we find a
destructive total interference at low energy in
the reaction $\pi^- p \to n e^+ e^-$,
while for $\pi^+n \to pe^+ e^-$ the interference is constructive.
In the $\rho$ channel also the $P_{11} (1440)$ resonance plays a role.
At higher energies (see Fig.~5)
for $\omega$ production the $S_{11}(1650)$ resonance is dominant
and, therefore, the total $\rho- \omega$ interference
for $\pi^-p \to ne^+ e^-$ ( $\pi^+n \to pe^+ e^-$) becomes constructive
(destructive) as depicted in Fig.~3.
For backward directions, the nucleon channel makes a noticeable contribution.

The relative contribution of different resonances depends on the energy,
and this dependence is exhibited in Fig.~6
for dominant resonances at $M=0.6$ GeV.
One can see the dominance of $S_{11}(1535)$ at low energy and
of $F_{15}(1680)$ at higher energies.
The  dominance of  $S_{11}(1535)$ at low energy
leads to a strong destructive (constructive) interference
in $\pi^-p\to ne^+e^-$ ( $\pi^+n\to pe^+e^-$) reactions shown in
Fig.~7, where we display the invariant mass distribution
as a function of energy at $M=0.6$ MeV.

In Fig.~8 we show the  energy dependence of the invariant mass distributions
at $M =$ 0.6 and 0.782 GeV.
One can see a striking difference for the two reactions
under consideration.
There is also a strong sensitivity on changes of the invariant 
dilepton mass $M$.

Fig.~9 (left panel) displays the energy dependence of the spin-density
matrix element $\rho_{00}$
\begin{eqnarray}
\rho_{00}=\frac{\Sigma_\parallel}{\Sigma_\parallel +2\Sigma_\perp},
\label{rho00}
\end{eqnarray}
at $\theta=30^o$. One can see a similar qualitative
behavior of $\rho_{00}$ for the two reactions. The corresponding angular
distributions of electrons, normalized to 1, are shown in Fig. 9 (right panel).
We have to note that near threshold $\rho_{00}$ is close to $\frac13$ which
results in an almost isotropic electron distribution. 
Far above the threshold,
for example at $s^{1/2}=1.8$ GeV and $M=0.6$ GeV,
the resonance $F_{15}(1680)$ becomes
dominant and $\rho_{00}$ exhibits an additional $\theta$ 
dependence with maxima at $\theta=0, \pi$ and a minimum
at $\theta=\frac\pi2$,  which leads to an anisotropy in the 
electron decay distributions.

\subsection{Adjusting couplings from resonance decays} 
\label{III.B}

All the above results are obtained with resonance parameters shown in
Table~1 and based on the quark model estimates in \cite{RB00}.
In \cite{LWF99,M_P_L} coupled channel calculations are performed
with the goal the extract the couplings from a combined analysis
of a large set of reaction data. 
To get an idea on the importance of a particular
set of coupling strengths within our approach
one should compare the above results with
such ones which rely on a different set.
In principle, one can try to get the absolute values of the $\rho NB^*$
coupling strengths by using the partial branching ratios of the decays
$B^* \to N \rho$ \cite{PDG98} via
\begin{eqnarray}
g_{\rho NB^*_i}^{\rm fit \quad 2} = \Gamma_{B^*_i\to N\rho}
\left[
\frac{2a_im_\rho{\Gamma_\rho}_0}{8\pi^2(2J_i+1) M_{B^*_i}^2}
\int_{2m_\pi}^{s^{1/2}-M_N} \frac{k(M)
F(M) M d M}{(M^2-m_\rho^2)^2 + (m_\rho\Gamma_\rho)^2}
\right]^{-1},
\end{eqnarray}
where $k(M)=\sqrt{M^2/4-m_\pi^2}$, $a_i = 3$ (1) for resonances with
isospin $\frac12$ ($\frac32$), and $J_i$ is the resonance spin. The
function $F(M)$ reads
\begin{eqnarray}
F(M)={\rm Sp}\left( (\not\hskip-0.4mm\!{p\,'} +M_N){\cal
G}^{\mu,\kappa} \, \Pi_{\kappa,\kappa'} \, {\cal G}^{\nu,\kappa'}
\right) (-g_{\mu\nu} + \frac{q_\mu g_\nu}{M^2}),
\end{eqnarray}
where ${\cal G}^{\mu,\kappa}$ ($\kappa=0,\alpha, \alpha\beta$)
according to Eqs.~(\ref{LomegaB}, \ref{LrhoB})
is taken from Eqs.~(\ref{NR1440} - \ref{NR1720}) and
\begin{eqnarray}
\Pi_{\kappa,\kappa'}=\sum_r {\cal U}^r_{\kappa}\bar{\cal U}^r_{\kappa'}
\end{eqnarray}
with ${\cal U}$ from Eq.~(\ref{R-S}).
The corresponding branching ratios and $g_{\rho NB^*}^{\rm fit}$
are shown in Table~2, where the phases are taken the same as
predicted by quark model \cite{RB00} shown in Table~1.
One can see that some of the couplings in Table~2 are smaller than
the corresponding values in Table~1
(cf.
$P_{11}(1440)$,
$D_{13}(1520)$,
$S_{11}(1535)$,
$S_{11}(1650)$,
$D_{13}(1700)$,
$D_{15}(1675)$,
$F_{15}(1680)$),
while for other resonances they are greater
(cf.
$P_{33}(1600)$,
$S_{31}(1620)$,
$D_{33}(1700)$,
$P_{13}(1720)$).
It should be emphasized that a calculation
with $g_{\rho NB^*}^{\rm fit}$ is not fully self consistent
because one can not fix the coupling $g_\omega NN^*$ by this method,
rather for them we use the prediction of the quark model \cite{RB00}
as listed in Table~1.
Nevertheless, for methodical purposes and to elucidate the sensitivity
of our results,
we perform such a calculation and present the results.

In Fig.~10 (left panel) we show the individual contributions of
resonances for the $\rho$ channel of the reaction
$\pi^+n \to p e^+ e^-$ as a function of $s^{1/2}$.
The resonances  $S_{11}(1535)$ and, somewhat less important, $P_{11}(1440)$
dominate at low energy,
while $P_{13}(1720)$ becomes stronger at higher energies;
in between $S_{11} (1650)$ is important.
The total contribution of the $\rho$ meson
with this new parameter set is smaller.
The differential cross sections are shown in  Fig.~10 (right panel).
The absolute value of the total cross section is smaller
than that shown in the left panel of Fig.~7.
But qualitatively their behavior is similar for both parameter sets.

In Fig.~11 we show the
differential cross section of dielectron production
for the reactions
$\pi^- p \to n e^+ e^-$ and $\pi^+ n \to p e^+ e^-$ as a function
of $s^{1/2}$ at $M=0.6$ GeV (left panel)
and as a function of $M$ at $s^{1/2}=1.8$ GeV (right panel).
The isospin effect is greater at low energy and low invariant mass,
as shown in the left panel.

In Fig.~12 we show the energy dependence of the spin-density
matrix element $\rho_{00}$ (left panel) and electron angular distribution
(right panel), as in Fig.~9. One can see a strong difference of $\rho_{00}$
for the two reactions and a deviation from  the results shown in Fig.~9:
a strong increase at low energy for the $\pi^ + n$ reaction,
because of a sizable contribution of the $P_{11}(1440)$ resonance, and
relatively small value of $\rho_{00}$ at higher energy, because of the
dominance of the $P_{13}(1720)$ resonance.
Also the angular distributions change considerably for this new
parameter set.
This sensitivity clearly demonstrates the need of experimental data
for constraining the parameter space.

\section{Summary} 

In summary we have performed a combined analysis of
the dielectron invariant mass distributions for the exclusive
reactions $\pi^- p \to n e^+ e^-$  and $\pi^+ n \to p e^+ e^-$ near
the $\omega$ threshold. The differential cross sections for
the two reactions are different because of
different $\rho^0 - \omega$ interferences.
The calculation is based on a resonance model
with $s$ and $u$ channels,
where the $\omega(\rho) NB^*$ couplings as well as
the phases of the $\pi NB^*$ couplings are either
taken from the recent work
\cite{RB00} or, at least partially, are determined from resonance decays.
The found isospin effect is sensitive to the resonance coupling parameters
and, therefore, may be used as a powerful tool for the study of the resonance
dynamics in dielectron production processes.

We have shown that our predictions can be experimentally
tested by measuring the angular distribution of decay particles
in reactions of the type $\pi N\to N V \to N e^+e^-$
which are accessible
with the pion beam at the HADES spectrometer at GSI/Darmstadt \cite{HADES}.
(Notice that for the inverse reactions with real photons
a sizeable isospin effect is found, see \cite{Nozawa}.) 
We propose for the first time a systematic study of the
isoscalar part of the electromagnetic current by using a combined
analysis of dielecton production in $\pi^+n$ and $\pi^-p$
reactions. To this end it would be desirable to have at our disposal
the ratio or the difference of the 
$\pi^+n$ and $\pi^-p$ cross sections 
(which might be deduced, e.g., from the reactions $\pi^+d$ and $\pi^-d$)
as a function of both
the invariant dilepton mass in the interval $M =0.6 \cdots 0.8$ GeV
and the energy in the interval 
$s^{1/2} = s^{1/2}_{\rm threshold} \cdots 1.9$ GeV.
This quantity is most sensible for a study of the 
$\rho-\omega$ interference.

Finally, it should be stressed that the present investigation is completely
based on the resonance model and, therefore, is valid near threshold.
At higher energy one has to include other mechanisms
like meson exchange $t$ channel amplitudes.
Unfortunately, in this case one has to make some assumptions
on the relative phase between $t$ and $s, u$ channels,
which is hitherto unknown.
However, we expect that the presented isospin effect will persist
in this case too.

\subsection*{Acknowledgments} 

We gratefully acknowledge  fruitful discussions with H.W. Barz,
R. Dressler, S.B. Gerasimov, J. Ritman, and Gy. Wolf.
In particular, B. Friman is thanked for a critical reading
of the manuscript.
One of the authors (A.I.T.) thanks for the warm hospitality
of the nuclear
theory group in the Research Center Rossendorf.
This work is supported by BMBF grant 06DR921
and Heisenberg-Landau program.

\appendix 

\section{Effective Lagrangians}

The effective Lagrangians for $\omega$ production within the framework of
the Riska-Brown model \cite{RB00} are listed in Ref.~\cite{TKR01}.
Therefore, here we focus on $\rho$ production
(the effective interaction Lagrangians needed for $\omega$ production
follow from the formulas below
by the substitution $\bbox{\rho} \to \omega$ and omitting corresponding
isospin factors, and skipping the $\Delta$ and $\Delta^*$ contributions).
For completeness we also include interactions with pions.
With the notation of subsection \ref{effective_L}
the relevant expressions read
\begin{eqnarray}
{\cal L}_{\pi, \rho NN}^{N_{\frac12^+}(940)\, N}
& = &
\bar\psi_N\,\left[ - \frac{f_{\pi NN}}{m_\pi}
\gamma_5\gamma_\mu\,\partial^\mu
\bbox{\pi \cdot \tau}\,\,
-\,\,{g_{\rho NN}}
\left(\gamma_\mu
- \frac{\kappa_{\rho NN}}{2M_N}
\sigma_{\mu\nu}\partial^\nu
\right)
\bbox{\rho}^\mu \bbox{\cdot \tau}\,
\right]\psi_{N},
\label{NR940}\\
{\cal L}_{\pi, \rho N\Delta}^{\Delta_{\frac32^+}(1232) P_{33}}
& = &
\bar\psi_N \, \left[
i\frac{f^{1232}_{\pi N\Delta}}{m_\pi}
\,\partial^\alpha
\bbox{\pi \cdot \chi}\,\,\right.
\nonumber\\
& & \left. \quad\quad - \frac{g^{1232}_{\rho NN^*}}{M_\Delta+M_N}
\,\gamma_5 \, \left( \gamma_\mu\partial^\alpha
- g_\mu^\alpha\,\not\hskip-0.7mm\!{\partial}
\right)
\bbox{\rho}^\mu \bbox{\cdot \chi}\,
\right]{\psi_{\Delta}}_{\alpha}\,\, + \,{\rm h.c.},
\label{NR1232}\\
{\cal L}_{\pi, \rho NN^*}^{N_{\frac12^+}(1440)\, P_{11}}
& = &
\bar\psi_N\, \left[ -\frac{f^{1440}_{\pi NN^*}}{m_\pi}
\gamma_5\gamma_\mu\,\partial^\mu
\bbox{\pi \cdot \tau}\,
\right.
\nonumber\\
& & \left. \quad\quad
-\,\,{g^{1440}_{\rho NN^*}}
\,(\gamma_\mu + \partial_\mu \not\hskip-0.7mm\!{\partial} \,m_\rho^{-2}
- \frac{\kappa_{\rho NN^*}}{M_{N^*}-M_N}
\sigma_{\mu\nu}\partial^\nu
)
\bbox{\rho}^\mu \bbox{\cdot \tau}\,
\right]\psi_{N^*} \,\, +  \,{\rm h.c.},
\label{NR1440}\\
{\cal L}_{\pi, \rho NN^*}^{N_{\frac32^-}(1520)\, D_{13}}
& = &
\bar\psi_N \, \left[
i \frac{f^{1520}_{\pi NN^*}}{m_\pi}
\,\gamma_5\,\partial^\alpha
\bbox{\pi \cdot \tau}\,\,
+\,\, \frac{g^{1520}_{\rho NN^*}}{m_\rho^2}
\sigma_{\mu\nu}\partial^\nu\partial^\alpha
\,
\bbox{\rho}^\mu \bbox{\cdot \tau}\,
\right]{\psi_{N^*}}_\alpha
\,\, + \,{\rm h.c.},
\label{NR1520}\\
{\cal L}_{\pi, \rho NN^*}^{N_{\frac12^-}(1535)\,S_{11}}
& = &
\bar\psi_N\, \left[
-\frac{f^{1535}_{\pi NN^*}}{m_\pi}
\gamma_\mu\,\partial^\mu
\bbox{\pi \cdot \tau}\,\,
-\,\,{g^{1535}_{\rho NN^*}}
\,\gamma_5
\,(\gamma_\mu + \partial_\mu\not\hskip-0.7mm\!{\partial}\, m_\rho^{-2} )
\bbox{\rho}^\mu \bbox{\cdot \tau}\,
\right]\psi_{N^*}
\,\, + \,{\rm h.c.},
\label{NR1535}\\
{\cal L}_{\pi, \rho N\Delta}^{\Delta_{\frac32^+}(1600) P_{33}}
& = &
\bar\psi_N \, \left[
i\frac{f^{1600}_{\pi N\Delta^*}}{m_\pi}
\,\partial^\alpha
\bbox{\pi \cdot \chi}\,\,\right.
\nonumber\\
& & \left. \quad\quad - \frac{g^{1600}_{\rho N\Delta^*}}{M_{\Delta^*} +M_N}
\,\gamma_5 \, \left( \gamma_\mu\partial^\alpha
- g_\mu^\alpha\,\not\hskip-0.7mm\!{\partial}
\right)
\bbox{\rho}^\mu \bbox{\cdot \chi}\,
\right]{\psi_{\Delta}}_{\alpha}\,\, + \,{\rm h.c.},
\label{NR1600}\\
{\cal L}_{\pi, \rho N\Delta^*}^{\Delta_{\frac12^-}(1620)\,S_{31}}
& = &
\bar\psi_N\, \left[
-\frac{f^{1620}_{\pi N\Delta^*}}{m_\pi}
\gamma_\mu\,\partial^\mu
\bbox{\pi \cdot \chi}\,\,
-\,\,{g^{1620}_{\rho N\Delta^*}}
\,\gamma_5
\,(\gamma_\mu + \partial_\mu\not\hskip-0.7mm\!{\partial}\, m_\rho^{-2} )
\bbox{\rho}^\mu \bbox{\cdot \chi}\,
\right]\psi_{\Delta}
\,\, + \,{\rm h.c.},
\label{NR1620}\\
{\cal L}_{\pi, \rho NN^*}^{N_{\frac12^-}(1650)\,S_{11}}
& = &
\bar\psi_N\, \left[
-\frac{f^{1650}_{\pi NN^*}}{m_\pi}
\gamma_\mu\,\partial^\mu
\bbox{\pi \cdot \tau}\,\,
-\,\,{g^{1650}_{\rho NN^*}}
\,\gamma_5
\,(\gamma_\mu + \partial_\mu\not\hskip-0.7mm\!{\partial}\, m_\rho^{-2} )
\bbox{\rho}^\mu \bbox{\cdot \tau}\,
\right]\psi_{N^*}
\,\, + \,{\rm h.c.},
\label{NR1650}\\
{\cal L}_{\pi, \rho NN^*}^{N_{\frac52^-}(1675)\,D_{15}}
& = &
\bar\psi_N\, \left[
-\frac{f^{1675}_{\pi NN^*}}{m_\pi^2}
\,\partial^\alpha\partial^\beta
\bbox{\pi \cdot \tau}\,\,
+\,\,\frac{g^{1675}_{\rho NN^*}}{m_\rho^2}
\,\epsilon^{\alpha \gamma \mu \nu}\,\gamma_\nu \,\partial_\gamma
\partial^\beta
\,\bbox{\rho}^\mu \bbox{\cdot \tau}\,
\right]
{\psi_{N^*}}_{\alpha\beta}
\,\, + \,{\rm h.c.},
\label{NR1675}\\
{\cal L}_{\pi, \rho NN^*}^{N_{\frac52^+}(1680) F_{15}}
& = &
\bar\psi_N\, \left[
-i\frac{f^{1680}_{\pi NN^*}}{m_\pi^2}
\,\gamma_5
\partial^\alpha\partial^\beta
\bbox{\pi \cdot \tau}\right.\,\,\nonumber\\
&& \left. \quad\quad
+\,\,\frac{g^{1680}_{\rho NN^*}}{m_\rho^2}
\,(\gamma_\mu + \partial_\mu\not\hskip-0.7mm\!{\partial}\, m_\rho^{-2} )
\,\partial^\alpha\partial^\beta
\,\bbox{\rho}^\mu \bbox{\cdot \tau}\,
\right]
{\psi_{N^*}}_{\alpha\beta}\,\, + \,{\rm h.c.},
\label{NR1680}\\
{\cal L}_{\pi, \rho N\Delta^*}^{\Delta_{\frac32^-}(1700) D_{33}}
& = &
\bar\psi_N\, \left[
i\frac{f^{1700}_{\pi N\Delta^*}}{m_\pi}
\,\gamma_5
\partial^\alpha
\bbox{\pi \cdot \chi}\,\,
+\,\,\frac{g^{1700}_{\rho N\Delta^*}}{m_\rho^2}\,
\sigma_{\mu\nu}\partial^\nu\partial^\alpha
\,\bbox{\rho}^\mu \bbox{\cdot \chi}\,
\right]{\psi_{\Delta}}_\alpha
\,\, + \,{\rm h.c.},
\label{NRD1700}\\
{\cal L}_{\pi, \rho NN^*}^{N_{\frac32^-}(1700) D_{13}}
& = &
\bar\psi_N\, \left[
i\frac{f^{1700}_{\pi NN^*}}{m_\pi}
\,\gamma_5
\partial^\alpha
\bbox{\pi \cdot \tau}\,\,
+\,\,\frac{g^{1700}_{\rho NN^*}}{m_\rho^2}\,
\sigma_{\mu\nu}\partial^\nu\partial^\alpha
\,\bbox{\rho}^\mu \bbox{\cdot \tau}\,
\right]{\psi_{N^*}}_\alpha
\,\, + \,{\rm h.c.},
\label{NR1700}\\
{\cal L}_{\pi, \rho NN^*}^{N_{\frac32^+}(1720) P_{13}}
& = &
\bar\psi_N \, \left[
i\frac{f^{1720}_{\pi NN^*}}{m_\pi}
\,\partial^\alpha
\bbox{\pi \cdot \tau}\,\,\right.
\nonumber\\
& & \left. \quad\quad - \frac{g^{1720}_{\rho NN^*}}{M_{N^*}+M_N}
\,\gamma_5 \, \left( \gamma_\mu\partial^\alpha
- g_\mu^\alpha\,\not\hskip-0.7mm\!{\partial}
\right)
\,\bbox{\rho}^\mu \bbox{\cdot \tau}\,
\right]{\psi_{N^*}}_{\alpha}\,\, + \,{\rm h.c.}.
\label{NR1720}
\end{eqnarray}
We use the convention of Bjorken and Drell \cite{BD}
in definitions of $\gamma$ matrices and the
spin matrix $\sigma_{\mu \nu}$.
The expressions Eqs.~(\ref{NR940} - \ref{NR1720})
are based on \cite{RB00}.

\section{Invariant amplitudes}

Here we list the  explicit expressions for the amplitudes
${\cal A}^\mu (N^*) \equiv 
{\cal A}^{(\rho) \mu} (N^*) =
{\cal A}^{(\rho) \mu}_s (N^*) -
{\cal A}^{(\rho) \mu}_u (N^*)$
(expressions for $\omega$ production follow
from them in a straightforward way) and
${\cal A}^\mu (\Delta^*) \equiv
{\cal A}^{(\rho) \mu} (\Delta^*) =
{\cal A}^{(\rho) \mu}_s (\Delta^*) +
{\cal A}^{(\rho) \mu}_u (\Delta^*)$
in Eq.~(\ref{T_rho_B*}),
\begin{eqnarray}
{\cal A}_\mu(N^{940} \, )
&=&
-i\frac{\Gamma^{(\rho)}_\mu (-q)\Lambda({p}_L,M_{N^*})
\gamma_5\not\hskip-0.7mm\!{k}\,
F_N(s)}{s-m_N^2}
+
i\frac{\gamma_5\not\hskip-0.7mm\!{k}\, \Lambda({p}_R,M_{N^*})
\Gamma^{(\rho)}_\mu (-q) F_N(u)}{u-m_N^2},
\label{AmplR1}\\
\nonumber\\
{\cal A}_\mu({\Delta^{1232}})
& = &
-i\frac{\gamma_5 \,(q^\alpha
\gamma_\mu - g_\mu^\alpha\not\hskip-0.7mm\!{q})\,
\Lambda_{\alpha\beta}({p}_L,M_{\Delta})k^\beta
F_{\Delta}(s)}{(M_{\Delta}+M_N)
(s-M_{\Delta}^2 +i\Gamma_{\Delta}M_{\Delta})},
\nonumber\\
&~& \hspace*{4cm}
- i \frac{k^\beta \Lambda_{\beta\alpha}({p}_R,M_{\Delta})
\gamma_5 \, (q^\alpha \gamma_\mu - g_\mu^\alpha\not\hskip-0.7mm\!{q})\,
F_{\Delta}(u)}{(M_{\Delta}+M_N)
(u-M_{\Delta}^2 + i\Gamma_{\Delta}M_{\Delta})},\\
\nonumber\\
{\cal A}_\mu({N^{1440}}) &=&
-i\frac{(\gamma_\mu
+ \frac{\kappa_{\rho NN^*}}{M_{N^*}-M_N}
q_\mu )
\Lambda({p}_L,M_{N^*})
\gamma_5\not\hskip-0.7mm\!{k}\,
F_{N^*}(s)}{s-M_{N^*}^2 +i\Gamma_{N^*}M_{N^*}}\nonumber\\
&&\quad\quad\quad\quad\quad\quad
+i\frac{\gamma_5\not\hskip-0.7mm\!{k}\,\Lambda({p}_R,M_{N^*})
(\gamma_\mu
+ \frac{\kappa_{\rho NN^*}}{M_{N^*}-M_N}
q_\mu )
F_{N^*}(u)}{u-M_{N^*}^2 + i\Gamma_{N^*}M_{N^*}},
\label{AmplNR1440}\\ \nonumber\\
{\cal A}_\mu({N^{1520}})
& = &
-\frac{\sigma_{\mu\nu}q^\nu q^\alpha
\Lambda_{\alpha\beta}({p}_L,M_{N^*})\gamma_5 k^\beta
F_{N^*}(s)}{m_\rho^2(s-M_{N^*}^2 +i\Gamma_{N^*}M_{N^*})}\nonumber\\
&~& \hspace*{4cm}
+ \frac{\gamma_5 k^\alpha\Lambda_{\alpha\beta}({p}_R,M_{N^*})
\sigma_{\mu\nu}\,q^{\nu}q^{\beta}
F_{N^*}(u)}{m_\rho^2(u-M_{N^*}^2 + i\Gamma_{N^*}M_{N^*})},\\
\nonumber\\
{\cal A}_\mu({N^{1535}}) &=&
-i\frac{\gamma_5\gamma_\mu\Lambda({p}_L,M_{N^*})
\not\hskip-0.7mm\!{k}\,
F_{N^*}(s)}{s-M_{N^*}^2 +i\Gamma_{N^*}M_{N^*}}
+
i\frac{\not\hskip-0.7mm\!{k}\,\Lambda({p}_R,M_{N^*})
\gamma_5\gamma_\mu F_{N^*}
(u)}{u-M_{N^*}^2 + i\Gamma_{N^*}M_{N^*}},\\
\nonumber\\
{\cal A}_\mu({\Delta^{1600}})
& = &
-i\frac{\gamma_5 \,(q^\alpha
\gamma_\mu - g_\mu^\alpha\not\hskip-0.7mm\!{q})\,
\Lambda_{\alpha\beta}({p}_L,M_{\Delta^*})k^\beta
F_{\Delta^*}(s)}{(M_{\Delta^*}+M_N)
(s-M_{\Delta^*}^2 +i\Gamma_{\Delta^*}M_{\Delta^*})},
\nonumber\\
&~& \hspace*{4cm}
- i \frac{k^\beta \Lambda_{\beta\alpha}({p}_R,M_{\Delta^*})
\gamma_5 \, (q^\alpha \gamma_\mu - g_\mu^\alpha\not\hskip-0.7mm\!{q})\,
F_{\Delta^*}(u)}{(M_{\Delta^*}+M_N)
(u-M_{\Delta^*}^2 + i\Gamma_{\Delta^*}M_{\Delta^*})},\\
\nonumber\\
{\cal A}_\mu({\Delta^{1620}})
& = &
-i\frac{\gamma_5\gamma_\mu\Lambda({p}_L,M_{\Delta^*})
\not\hskip-0.7mm\!{k}\,
F_{\Delta^*}(s)}{s-M_{\Delta^*}^2 +i\Gamma_{\Delta^*}M_{N\Delta^*}}
-
i\frac{\not\hskip-0.7mm\!{k}\,\Lambda({p}_R,M_{\Delta^*})
\gamma_5\gamma_\mu F_\Delta^*(u)}
{u-M_{\Delta^*}^2 + i\Gamma_{\Delta^*}M_{\Delta^*}},\\
\nonumber\\
{\cal A}_\mu({N^{1650}})
& = &
-i\frac{\gamma_5\gamma_\mu\Lambda({p}_L,M_{N^*})
\not\hskip-0.7mm\!{k}\,
F_{N^*}(s)}{s-M_{N^*}^2 +i\Gamma_{N^*}M_{N^*}}
+
i\frac{\not\hskip-0.7mm\!{k}\,\Lambda({p}_R,M_{N^*})
\gamma_5\gamma_\mu F_{N^*}(u)}{u-M_{N^*}^2 + i\Gamma_{N^*}M_{N^*}},\\
 \nonumber\\
{\cal A}_\mu({N^{1675}})
& = &
-\frac{\epsilon^\alpha_{\tau \mu \nu} q^\tau q^\beta k^\gamma k^\delta }
{m_\pi m_\rho^2}
\left(
\frac{\gamma^\nu
\Lambda_{\alpha \beta, \gamma \delta}({p}_L,M_{N^*})
F_{N^*}(s)}{s-M_{N^*}^2 +i\Gamma_{N^*}M_{N^*}}\right.\nonumber\\
&~& \hspace*{4cm}
- \left.
\frac{\Lambda_{\gamma \delta, \alpha \beta}({p}_R,M_{N^*}) \gamma^\nu
F_{N^*}(u)}{u-M_{N^*}^2 + i\Gamma_{N^*}M_{N^*}}
\right),\\  \nonumber\\
{\cal A}_\mu({N^{1680}})
& = &
-i \frac{q^\alpha q^\beta k^\gamma k^\delta }
{m_\pi m_\rho^2}
\left(
\frac{\gamma_\mu
\Lambda_{\alpha \beta, \gamma \delta}({p}_L,M_{N^*})\,\gamma_5
F_{N^*}(s)}{s-M_{N^*}^2 +i\Gamma_{N^*}M_{N^*}}\right.\nonumber\\
&~& \hspace*{4cm}
- \left.
\frac{\gamma_5\Lambda_{\gamma \delta, \alpha \beta}({p}_R,M_{N^*})\,
F_{N^*}(u)}{u-M_{N^*}^2 + i\Gamma_{N^*}M_{N^*}}
\right),
\nonumber\\
{\cal A}_\mu({\Delta^{1700}})
& = &
-\frac{\sigma_{\mu\nu}q^\nu q^\alpha
\Lambda_{\alpha\beta}({p}_L,M_{\Delta^*})\,\gamma_5 k^\beta
F_{\Delta^*}(s)}
{m_\rho^2
(s-M_{\Delta^*}^2 +i\Gamma_{\Delta^*}M_{\Delta^*})}\nonumber\\
&~& \hspace*{4cm}
- \frac{\gamma_5 k^\alpha\Lambda_{\alpha\beta}({p}_R,M_{\Delta^*})
\sigma_{\mu\nu} q^{\nu}q^{\beta}
F_{\Delta^*}(u)}
{m_\rho^2
(u-M_{\Delta^*}^2 + i\Gamma_{\Delta^*}M_{\Delta^*})},\\
\nonumber\\
{\cal A}_\mu({N^{1700}})
& = &
-\frac{\sigma_{\mu\nu}q^\nu q^\alpha
\Lambda_{\alpha\beta}({p}_L,M_{N^*})\,\gamma_5 k^\beta
F_{N^*}(s)}{m_\rho^2(s-M_{N^*}^2 +i\Gamma_{N^*}M_{N^*})}\nonumber\\
&~& \hspace*{4cm}
+ \frac{\gamma_5 k^\alpha\Lambda_{\alpha\beta}({p}_R,M_{N^*})
\sigma_{\mu\nu} q^{\nu}q^{\beta}
F_{N^*}(u)}{m_\rho^2(u-M_{N^*}^2 + i\Gamma_{N^*}M_{N^*})},\\
\nonumber\\
{\cal A}_\mu({N^{1720}})
& = &
-i\frac{\gamma_5 \,
(q^\alpha \gamma_\mu - g_\mu^\alpha\not\hskip-0.7mm\!{q})\,
\Lambda_{\alpha\beta}({p}_L,M_{N^*})k^\beta
F_{N^*}(s)}
{(M_{N^*}+M_N)(s-M_{N^*}^2 +i\Gamma_{N^*}M_{N^*})},
\nonumber\\
&~& \hspace*{4cm}
+ i \frac{k^\beta \Lambda_{\beta\alpha}({p}_R,M_{N^*})
\gamma_5 \, (q^\alpha \gamma_\mu - g_\mu^\alpha\not\hskip-0.7mm\!{q})\,
F_{N^*}(u)}
{(M_{N^*}+M_N)(u-M_{N^*}^2 + i\Gamma_{N^*}M_{N^*})},
\label{AmplAR}
\end{eqnarray}
with $p_L=p+k$, $p_R=p-q$ and
\begin{eqnarray}
\Gamma^{(\rho)}_\alpha (k_{(\rho)})
=
\gamma_\alpha
 + i\frac{\kappa_{\rho NN}}{2M_N} \sigma_{\alpha \beta} \,
k_{( \rho )}^{\beta}.
\label{Gamma}
\end{eqnarray}

For completeness we display also expressions for propagators
and Rarita-Schwinger spinors.
The resonance propagators in Eqs.~(\ref{AmplR1} - \ref{AmplAR})
are defined by the conventional  method \cite{IZ}
assuming the validity of the spectral decomposition
\begin{eqnarray}
\psi_{N^*}(x)=\int \frac{d^3{\bf p} }{(2\pi)^3\sqrt{2E_p}}
\left[
a_{{\bf p},r} \, u_{N^*}^r(p)e^{-ipx} +
b^{+}_{{\bf p},r} \, v_{N^*}^r(p)e^{+ipx}
\right].
\end{eqnarray}
The finite decay width $\Gamma_{N^*}$ is introduced into
the propagator denominators by substituting
$M_{N^*} \to M_{N^*}  - \frac i2 \Gamma_{N^*}$.
Therefore, the operators $\Lambda_{\cdots}({p},M)$ are defined as
\begin{eqnarray}
\Lambda({p},M)
& = &
\frac12 \sum_r \left(
(1+\frac{p_0}{E_0})u^r({\bf p},E_0)\otimes\bar u^r({\bf p},E_0)
\right. \nonumber \\
& &  \left. \hspace*{10.5mm}
-(1-\frac{p_0}{E_0})v^r({-\bf p},E_0)\otimes\bar v^r({-\bf p},E_0)
\right)
= \, \not\hskip-0.7mm\!{p}\,+\,M,
\label{LR-S} \\
\Lambda_{\alpha\beta}({p},M)
& = &
\frac12 \sum_r \left(
(1+\frac{p_0}{E_0}){\cal U}^r_\alpha({\bf p},E_0)
\otimes\bar {\cal U}^r_{\beta}({\bf p},E_0)
\right. \nonumber \\
& &  \left. \hspace*{10.5mm}
-(1-\frac{p_0}{E_0}){\cal V}^r_\alpha({-\bf p},E_0)
\otimes\bar {\cal V}^r_\beta({-\bf p},E_0)
\right), \\
\Lambda_{\alpha\beta,\gamma\delta}({p},M)
& = &
\frac12 \sum_r \left(
(1+\frac{p_0}{E_0}){\cal U}^r_{\alpha\beta}({\bf p},E_0)
\otimes\bar {\cal U}^r_{\gamma\delta}({\bf p},E_0)
\right. \nonumber \\
& & \left. \hspace*{10.5mm}
-(1-\frac{p_0}{E_0}){\cal V}^r_{\alpha\beta}({-\bf p},E_0)
\otimes\bar {\cal V}^r_{\gamma\delta}({-\bf p},E_0)
\right),
\end{eqnarray}
where $E_0=\sqrt{{\bf p}^2 + M^2}$, and the Rarita-Schwinger spinors read
\begin{eqnarray}
{\cal U}^r_\alpha(p)
& = &
\sum_{\lambda, s} \langle 1\, \lambda \,\frac12\,s|\,\frac32\,
r \rangle \, \varepsilon^\lambda_\alpha(p) \, u^s(p),\\
{\cal U}^r_{\alpha\beta}(p)
& = &
\sum_{\lambda, \lambda' s, t} \langle 1
\, \lambda \,\frac12 \, s| \, \frac32 \, t \rangle \,
\langle \frac32 \, t \, 1\, \lambda' | \, \frac52\, r \rangle\,
\varepsilon^\lambda_\alpha(p) \,
\varepsilon^{\lambda'}_\beta(p) \,
u^s(p).
\label{R-S}
\end{eqnarray}
The spinors $v$ and ${\cal V}$ are related to $u$ and ${\cal U}$ as
$v(p) = i \gamma_2\, u^*(p)$ and
${\cal V}(p)=i\gamma_2\, {\cal U}^*(p)$, respectively.
In our calculations we use energy-dependent total resonance
decay widths $\Gamma_{N^*}$. However, taking into account that the
effect of a finite width is quite different for $s$ and $u$ channels,
because of the evident relation
$|u|+M^2_{N^*} \gg |s-M^2_{N^*}|$,
we use $\Gamma_{N^*} = \Gamma^0_{N^*}$ for the $u$ channels and
\begin{equation}
\Gamma_{N^*} = \Gamma^0_{N^*}
\left[ 1 - B^\pi_{N^*} + B^\pi_{N^*}\,
\left( \frac{\bf k_{}}{{\bf k}_0} \right)^{2J} \right]
\end{equation}
for the $s$ channels, where $\Gamma^0_{N^*}$ is the total on-shell resonance
decay width and $B^\pi_{N^*}$ stands for the branching ratio
of the $N^*\to N\pi$ decay channel taken from \cite{PDG98};
${\bf k}^{}_0$ is the pion momentum at the resonance position,
i.e.  at $\sqrt{s}=M_{N^*}$,
and the factor
$({\bf k}/{{\bf k}_0})^{2J}$ comes from a direct
calculation of the $N^*\to N\pi$ decay width using the effective Lagrangians
of Eqs.~(\ref{NR1440} - \ref{NR1720}),
where we keep the leading term proportional to ${\bf k}^{2J}$.

\begin{table} 
\centering
\begin{tabular}{ccccccc}

baryon & $M_{B^*} $ & $f_{\pi NB^*}$ & $g_{\omega NB^*}$ & $g_{\rho NB^*}$ &
$\Gamma^0_{B^*} $ &$B^{\pi}_{B^*} $  \\ \hline
$N\frac12^+\,N     $ & $940$  & $1.0$     & $10.35$ & $3.0 $  & ---    & ---   \\ \hline
$N\frac12^+\,P_{11}$ & $1440$ & $0.39$    & $6.34$  & $1.78$  &$350$   &$0.65$\\ \hline
$N\frac32^-\,D_{13}$ & $1520$ & $-1.56$   & $8.88$  & $5.0 $  &$120$   &$0.55$\\ \hline
$N\frac12^-\,S_{11}$ & $1535$ & $0.36$    & $-5.12$ & $-2.9$  &$150$   &$0.45$\\ \hline
$N\frac12^-\,S_{11}$ & $1650$ & $0.31$    & $2.56$  & $-0.72$ &$150$   &$0.73$\\ \hline
$N\frac52^-\,D_{15}$ & $1675$ & $0.10$    & $10.87$ & $-3.1$  &$150$   &$0.45$\\ \hline
$N\frac52^+\,F_{15}$ & $1680$ & $-0.42$  & $-14.07$ & $-19.8$ &$130$   &$0.65$\\ \hline
$N\frac32^-\,D_{13}$ & $1700$ & $0.36$    & $2.81$  & $-0.45$  &$100$  &$0.10$\\ \hline
$N\frac32^+\,P_{13}$ & $1720$ & $-0.25$   & $-3.17$ & $-4.46$  &$150$  &$0.15$\\ \hline
$\Delta\frac32^+\,P_{33}$ & $1232$ & $2.21$    & $-    $ & $17.32$  &$120$   &$0.99$\\ \hline
$\Delta\frac32^+\,P_{33}$ & $1600$ & $0.52$    & $-    $ & $17.1 $  &$350$   &$0.18$\\ \hline
$\Delta\frac12^-\,S_{31}$ & $1620$ & $-0.17$   & $-    $  & $0.88$  &$150$   &$0.25$\\ \hline
$\Delta\frac32^-\,D_{33}$ & $1700$ & $1.32$    & $-    $  & $1.53$  &$300$   &$0.15$ \\
\end{tabular}
\vspace{0.5cm}
\caption{
Parameters for the resonance masses, coupling constants,
total decay widths and branching ratios for $N^*\to N\pi$ decays.
The resonance masses and decay widths are in units of MeV.}
\label{tab:Nstar1}
\end{table}

\begin{table} 
\centering
\begin{tabular}{cccc}
baryon & $M_{B^*} $  & $g_{\rho NB^*}^{\rm fit}$  &$B^{\rho}_{B^*} $  \\ \hline
$N\frac12^+\,P_{11}$ & $1440$ & $1.07$      &$<0.08$\\ \hline
$N\frac32^-\,D_{13}$ & $1520$ & $2.70$      &$0.15-0.25$\\ \hline
$N\frac12^-\,S_{11}$ & $1535$ & $-0.63$     &$<0.04$\\ \hline
$N\frac12^-\,S_{11}$ & $1650$ & $-0.49$     &$0.04-0.12$\\ \hline
$N\frac52^-\,D_{15}$ & $1675$ & $-0.79$     &$0.01-0.03$\\ \hline
$N\frac52^+\,F_{15}$ & $1680$ & $-1.19$     &$0.03-0.15$\\ \hline
$N\frac32^-\,D_{13}$ & $1700$ & $-1.31$     &$<0.35$\\ \hline
$N\frac32^+\,P_{13}$ & $1720$ & $-13.9$     &$0.7-0.85$\\ \hline
$\Delta\frac32^+\,P_{33}$ & $1600$ & $41.0$      &$<0.25$\\ \hline
$\Delta\frac12^-\,S_{31}$ & $1620$ & $1.39$      &$0.07-0.25$\\ \hline
$\Delta\frac32^-\,D_{33}$ & $1700$ & $4.27$      &$0.30-0.55$ \\
\end{tabular}
\vspace{0.5cm}
\caption{
Parameters for the  coupling constants $g_{\rho NN^*}^{\rm fit}$
calculated from the partial decay widths
$\Gamma_{B^* \to N \rho}$ \protect\cite{PDG98}.}
\label{tab:Nstar2}
\end{table}
\begin{figure}
\centering
{\epsfig{file=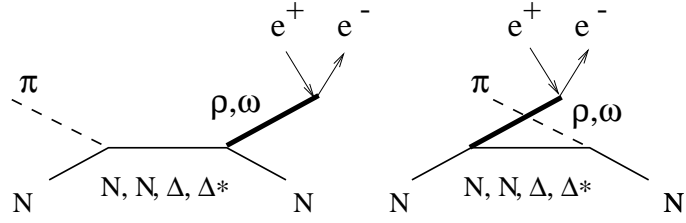, width=9cm}}
\bigskip
\caption{
Diagrammatic representation of the reaction 
$\pi N \to N e^+ e^-$ 
for the $s$ and $u$ channels.}
\label{fig:1}
\end{figure}

\begin{figure}
\centering
{\epsfig{file=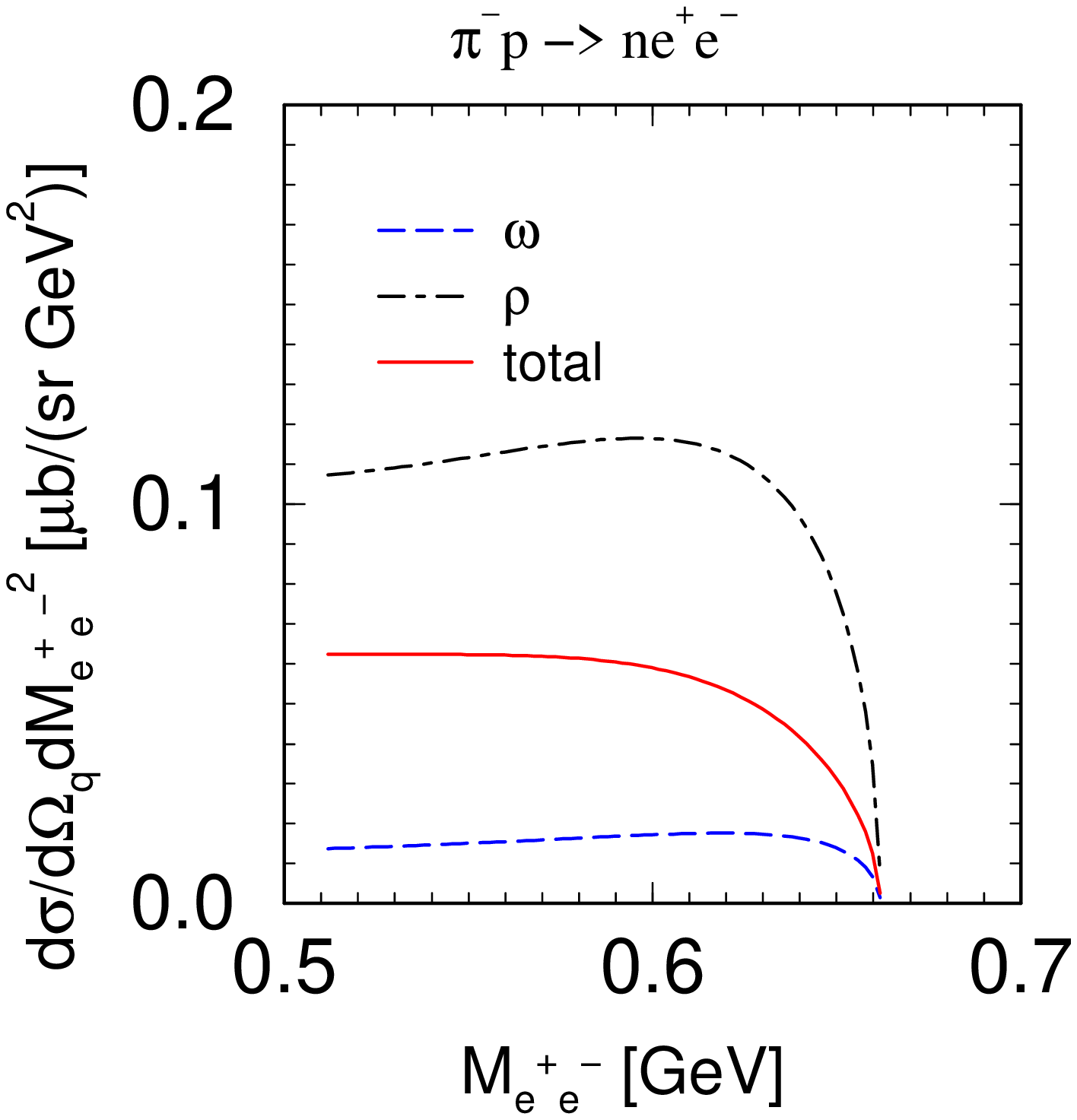 , width=6.90cm}\qquad\qquad
 \epsfig{file=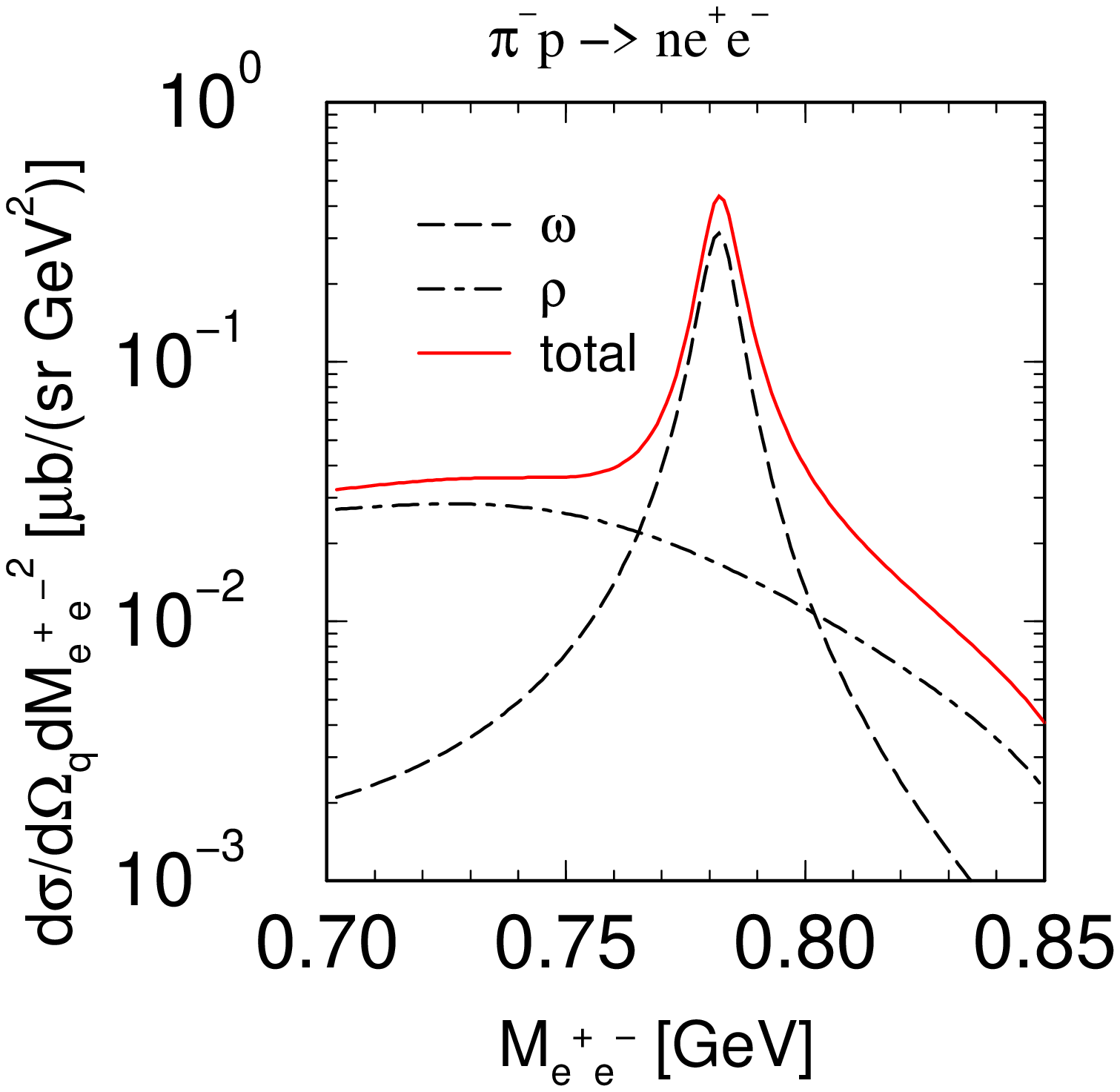 , width=7.30cm}}
\caption{
Differential cross sections of dielectron production
for the reactions
$\pi^- p \to n e^+e^-$ as a function of the dielectron invariant mass
for $s^{1/2}=1.6$ GeV (left panel) and
$s^{1/2}=1.8$ GeV (right panel). Dashed and dot-dashed
lines correspond to separate $\omega$ and $\rho$ contributions,
while solid lines are for the coherent sums.}
\label{fig:2}
\end{figure}

\begin{figure}
\centering
{\epsfig{file=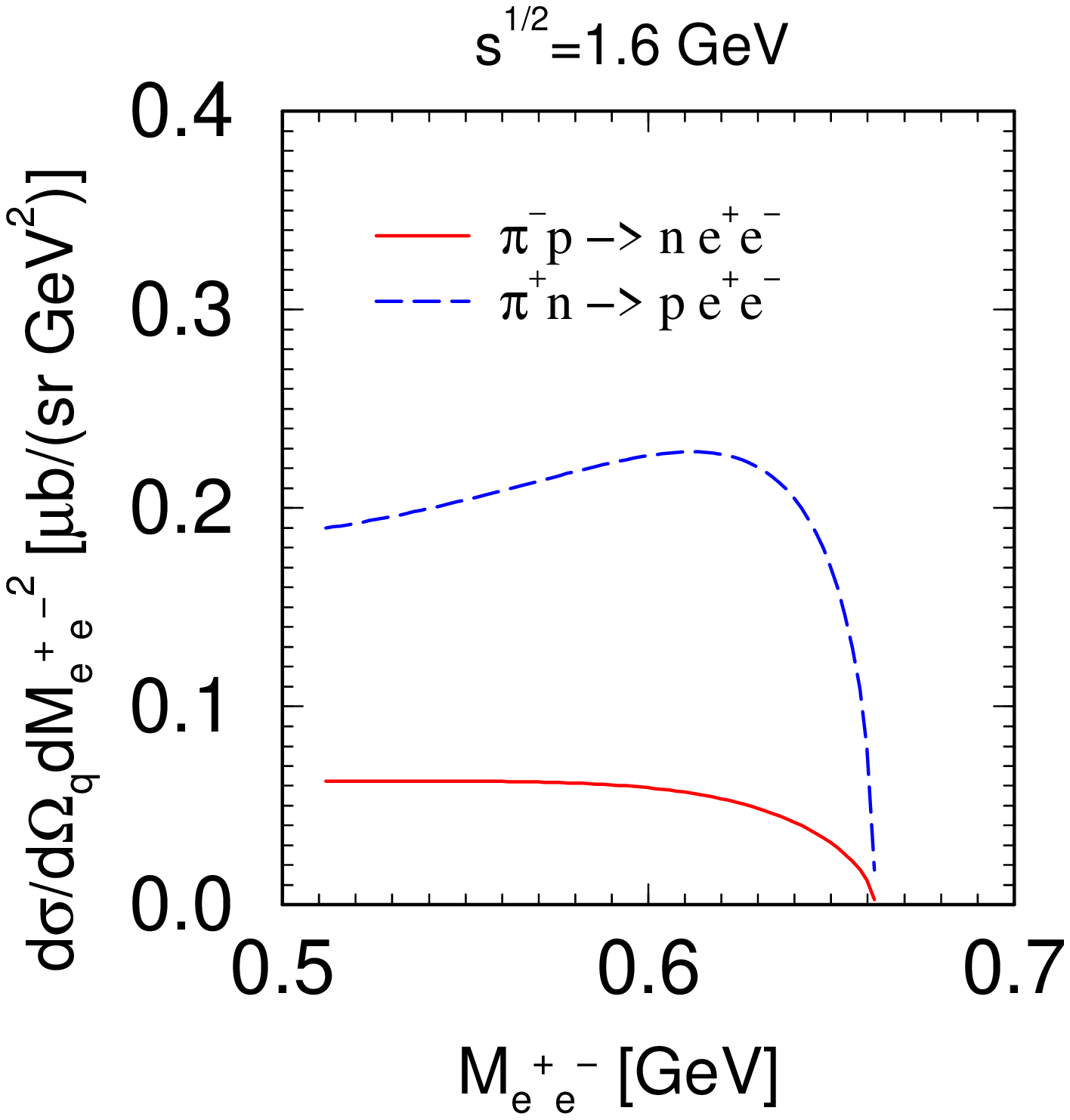 , width=6.90cm}\qquad\qquad
 \epsfig{file=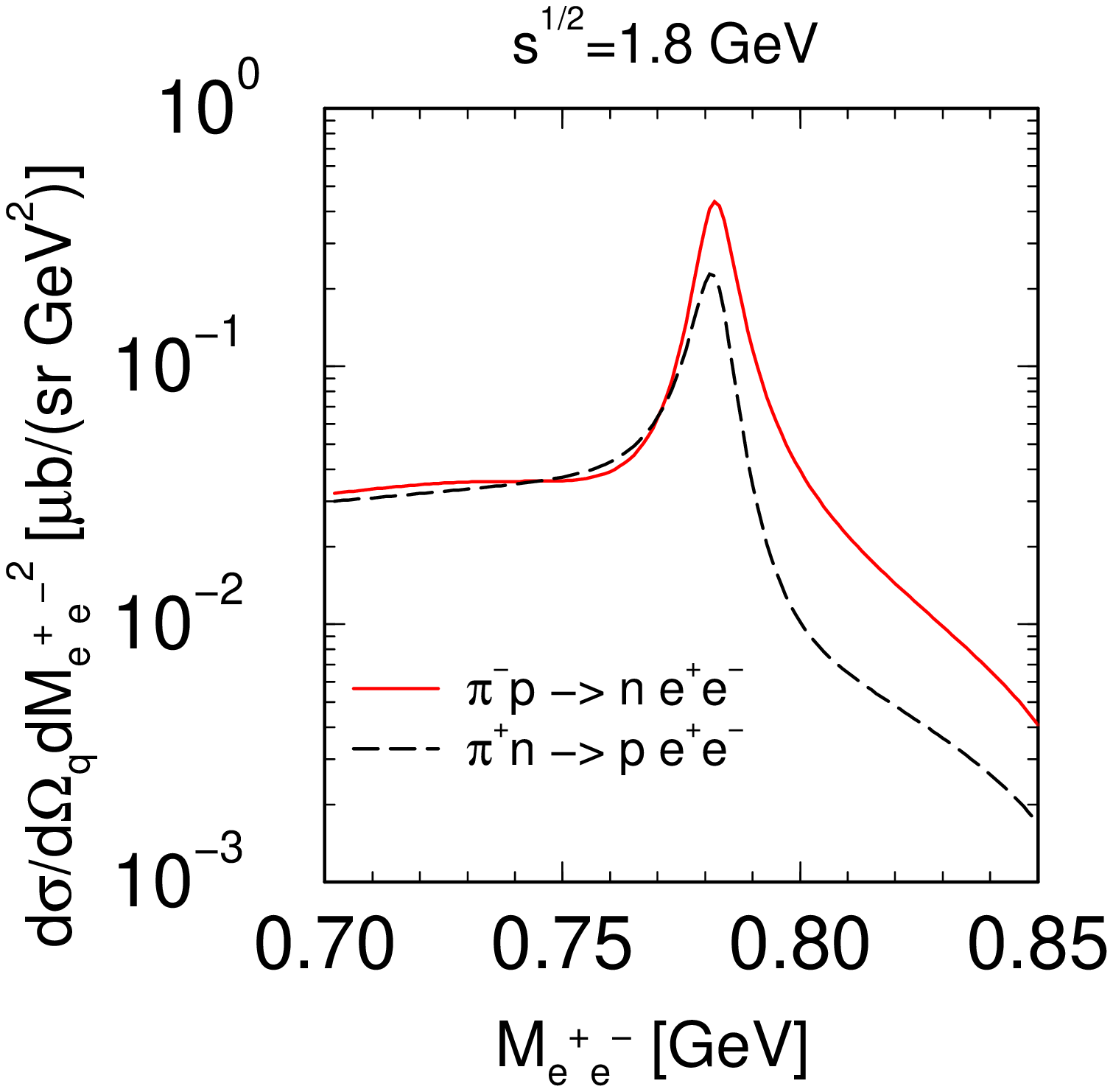 , width=7.30cm}}
\caption{
Differential cross sections of dielectron production
for the reactions
$\pi^- p\to n e^+e^-$ (solid lines) and $\pi^+ n\to p e^+e^-$
(dashed lines) as a function of dielectron invariant mass
for $s^{1/2}=1.6$ GeV (left panel) and 1.8 GeV (right panel). }
\label{fig:3}
\end{figure}

\begin{figure}
\centering
{\epsfig{file=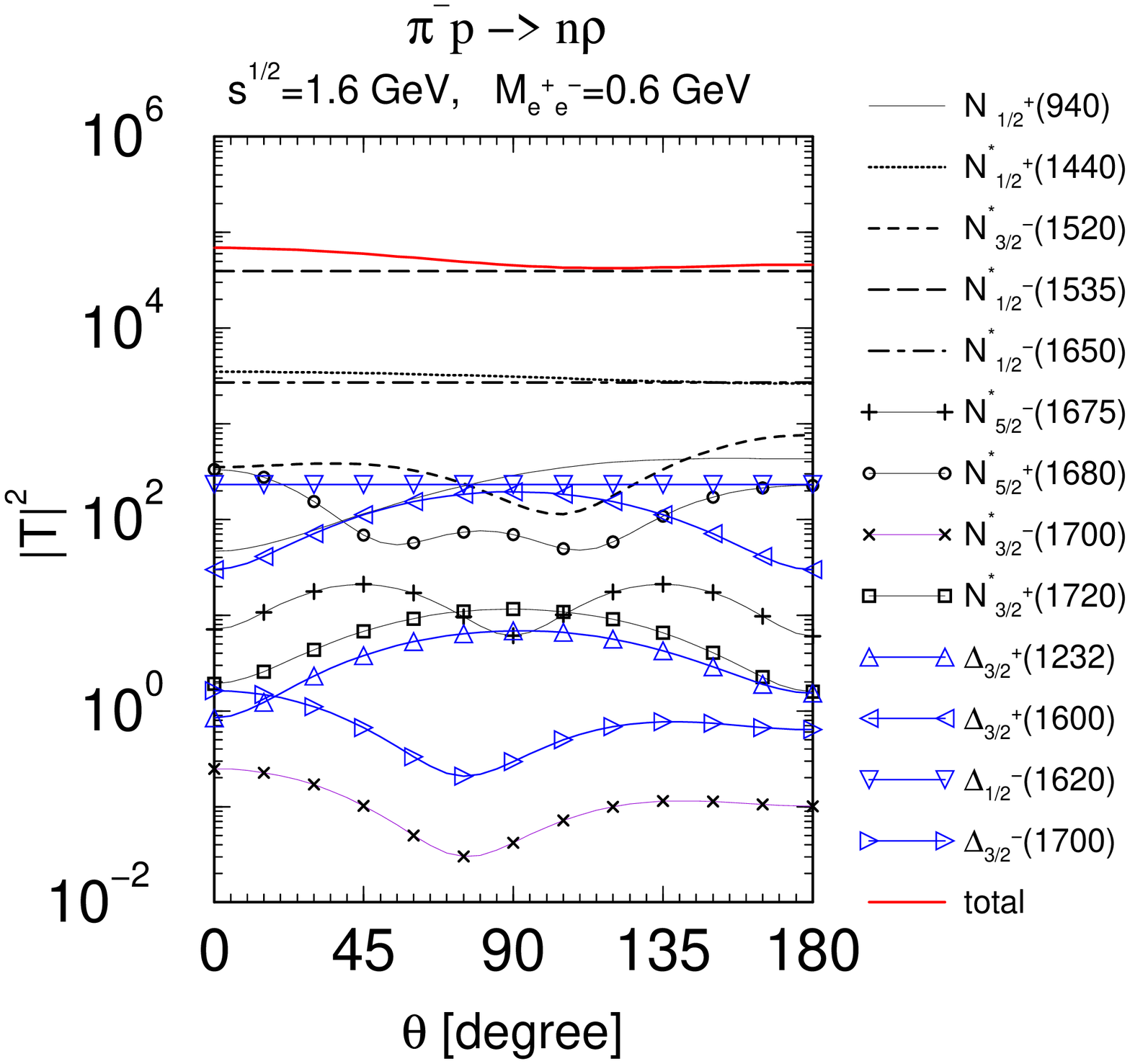, width=7.7cm}\qquad\qquad
\epsfig{file=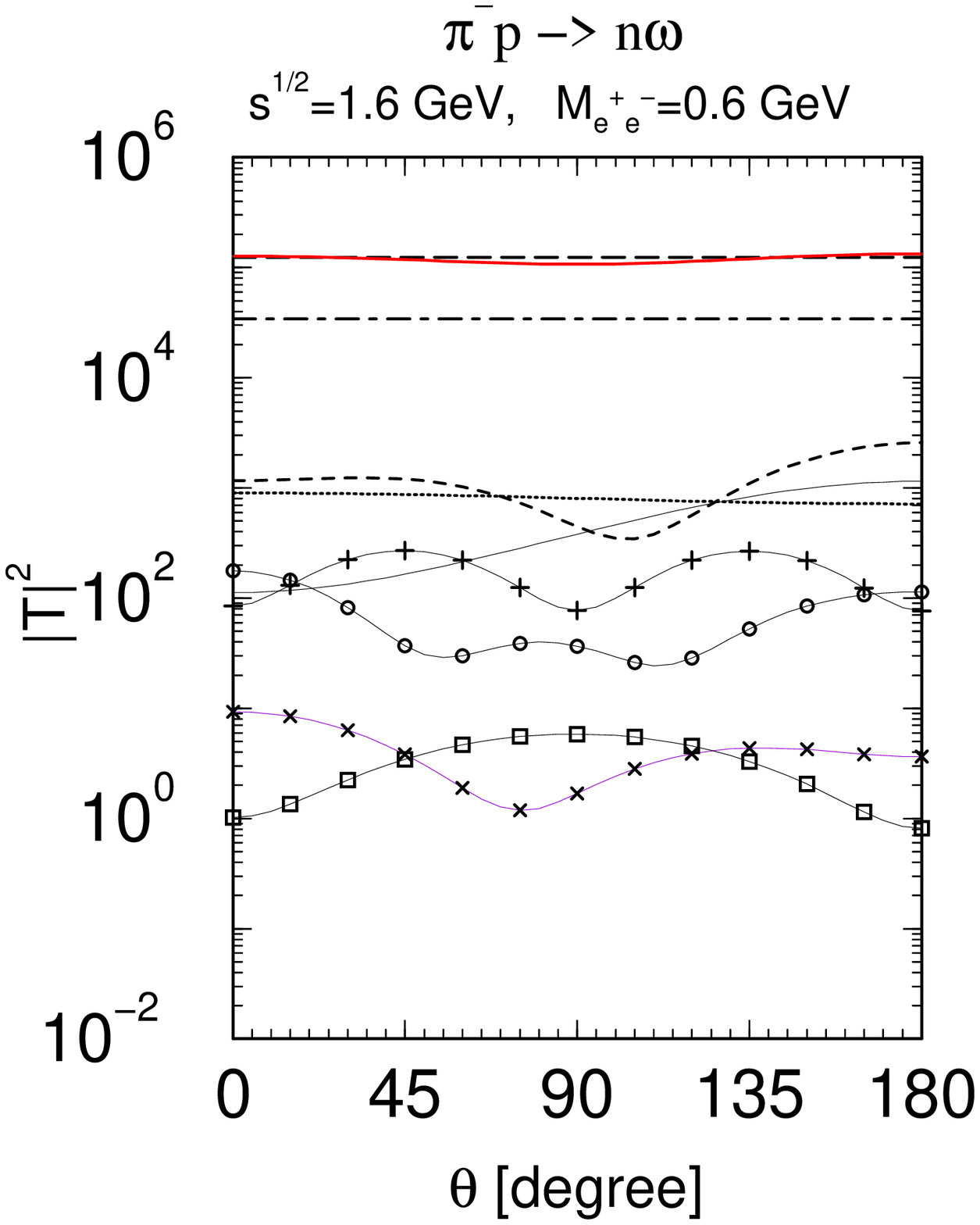, width=6.3cm}}
\caption{
Angular distribution of the
individual contributions of nucleon resonances listed in Table~1
to the spin averaged invariant amplitude
of $\rho$ (left panel)
and $\omega$ (right panel) channels
at  $s^{1/2} = 1.6$ GeV and for $M_{e^+e^-}=0.6$ GeV.}
\label{fig:4}
\end{figure}

\begin{figure}
\centering
{\epsfig{file=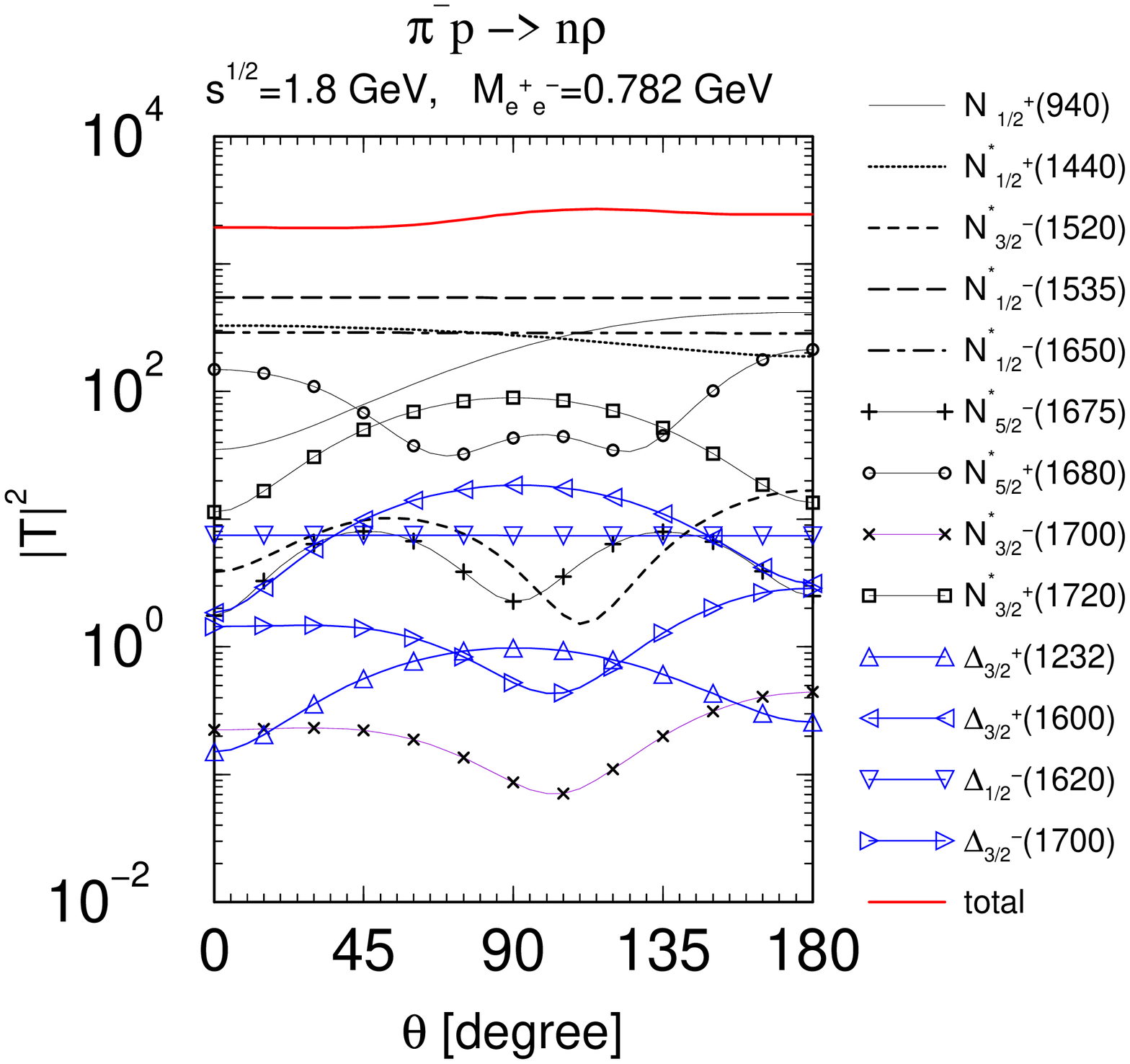, width=7.7cm}\qquad\qquad
 \epsfig{file=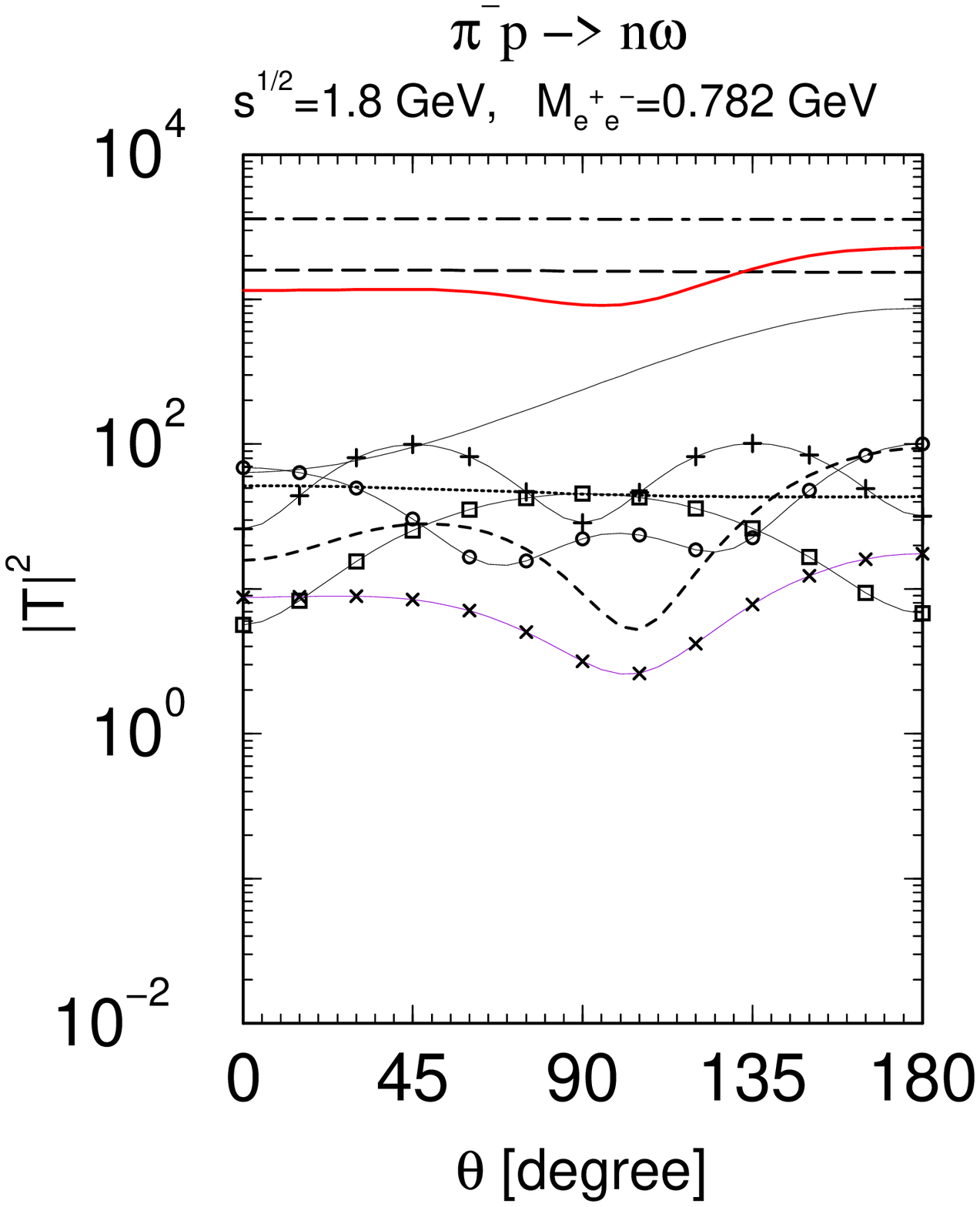, width=6.4cm}}
\caption{
The same as in Fig.~4 but at  $s^{1/2} = 1.8$ MeV and $M_{e^+e^-}=0.782$ GeV.}
\label{fig:5}
\end{figure}

\begin{figure}
\centering
{\epsfig{file=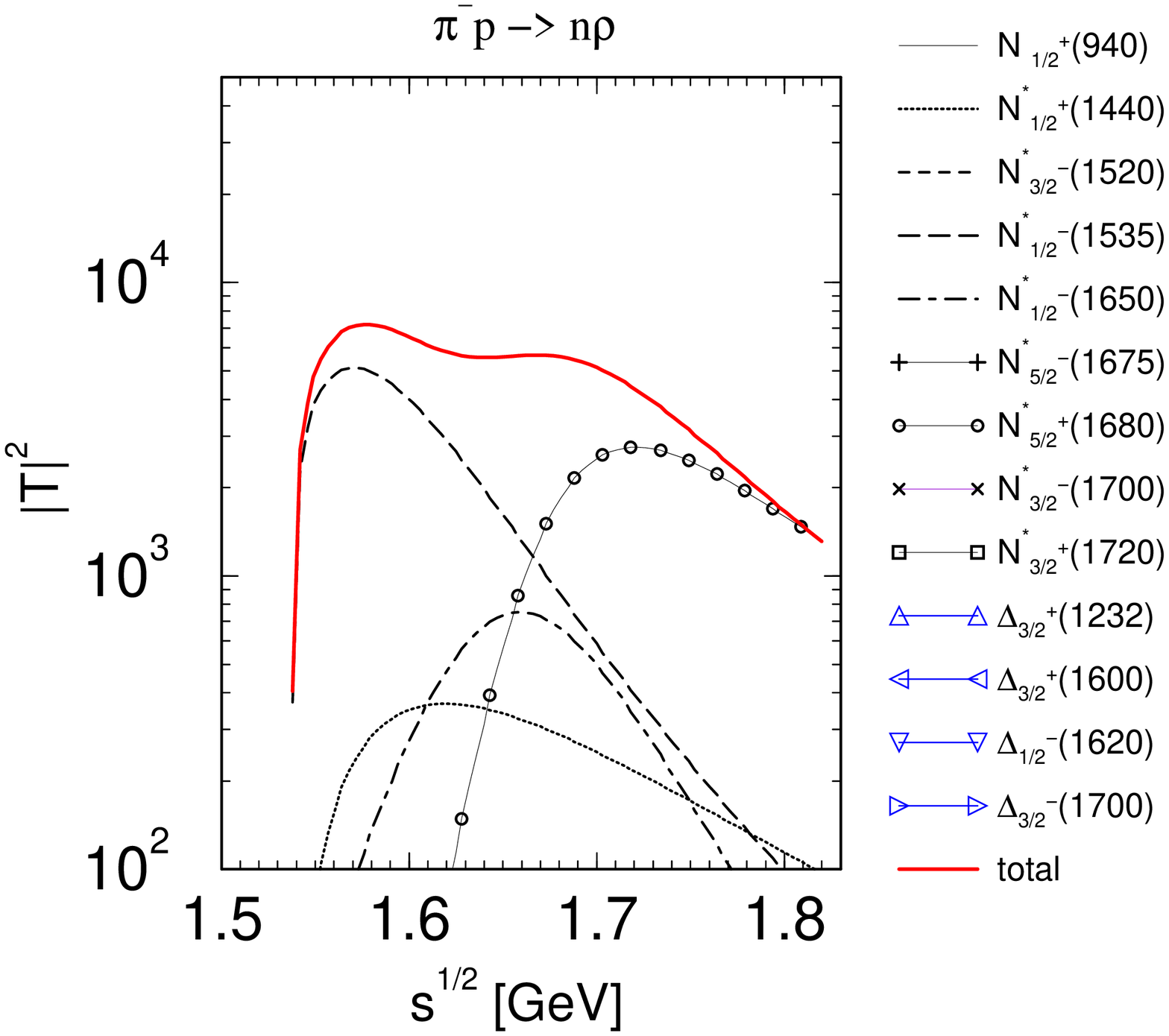, width=7.8cm}\qquad\qquad
\epsfig{file=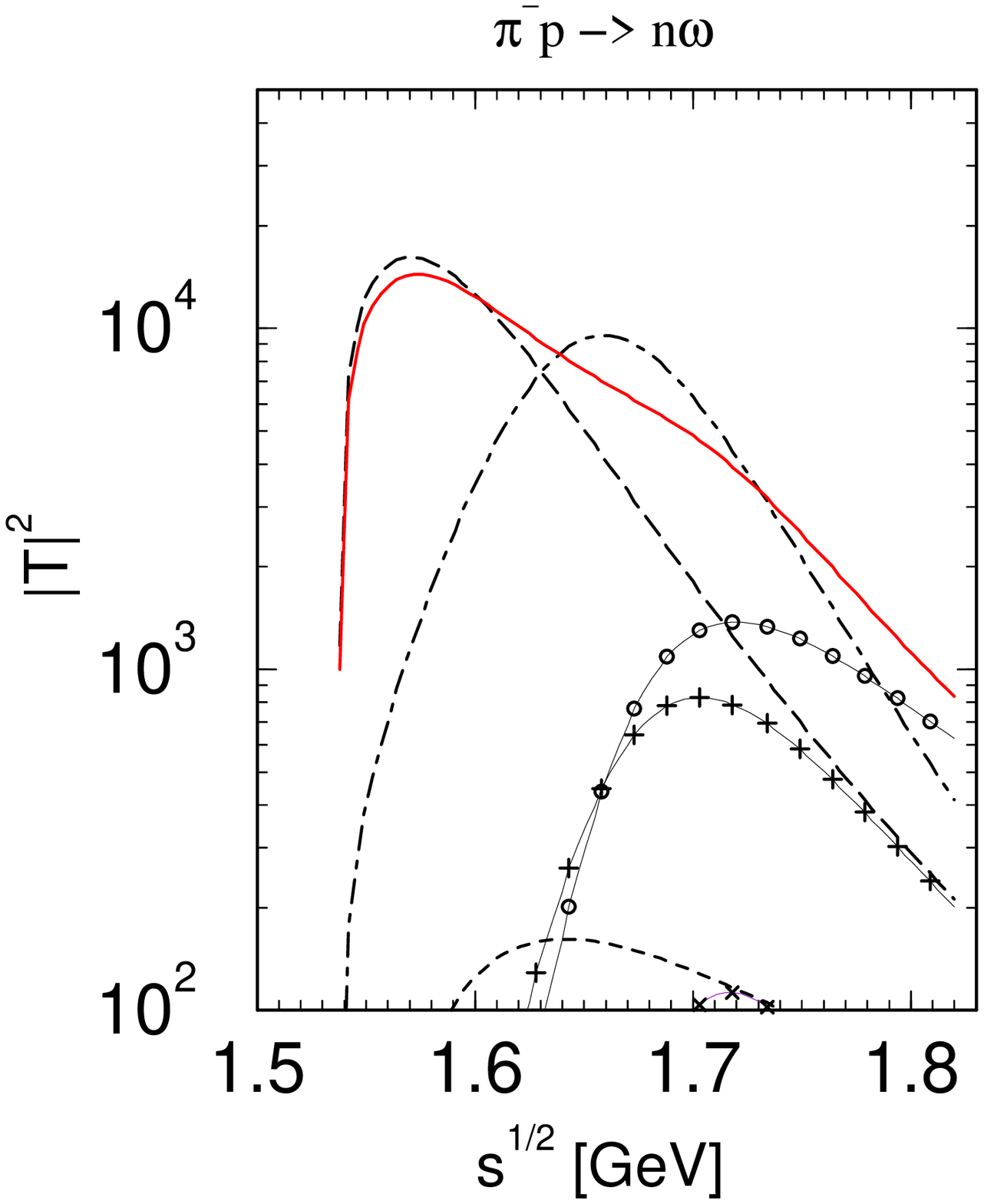, width=6.1cm}}
\caption{
Individual contributions of nucleon resonances
to the spin averaged invariant amplitude
of $\rho$ (left panel)
and $\omega$ (right panel) channels
as a function of
$s^{1/2}$ at $M_{e^+e^-}=0.6$ GeV.}
\label{fig:6}
\end{figure}

\begin{figure}
\centering
{\epsfig{file=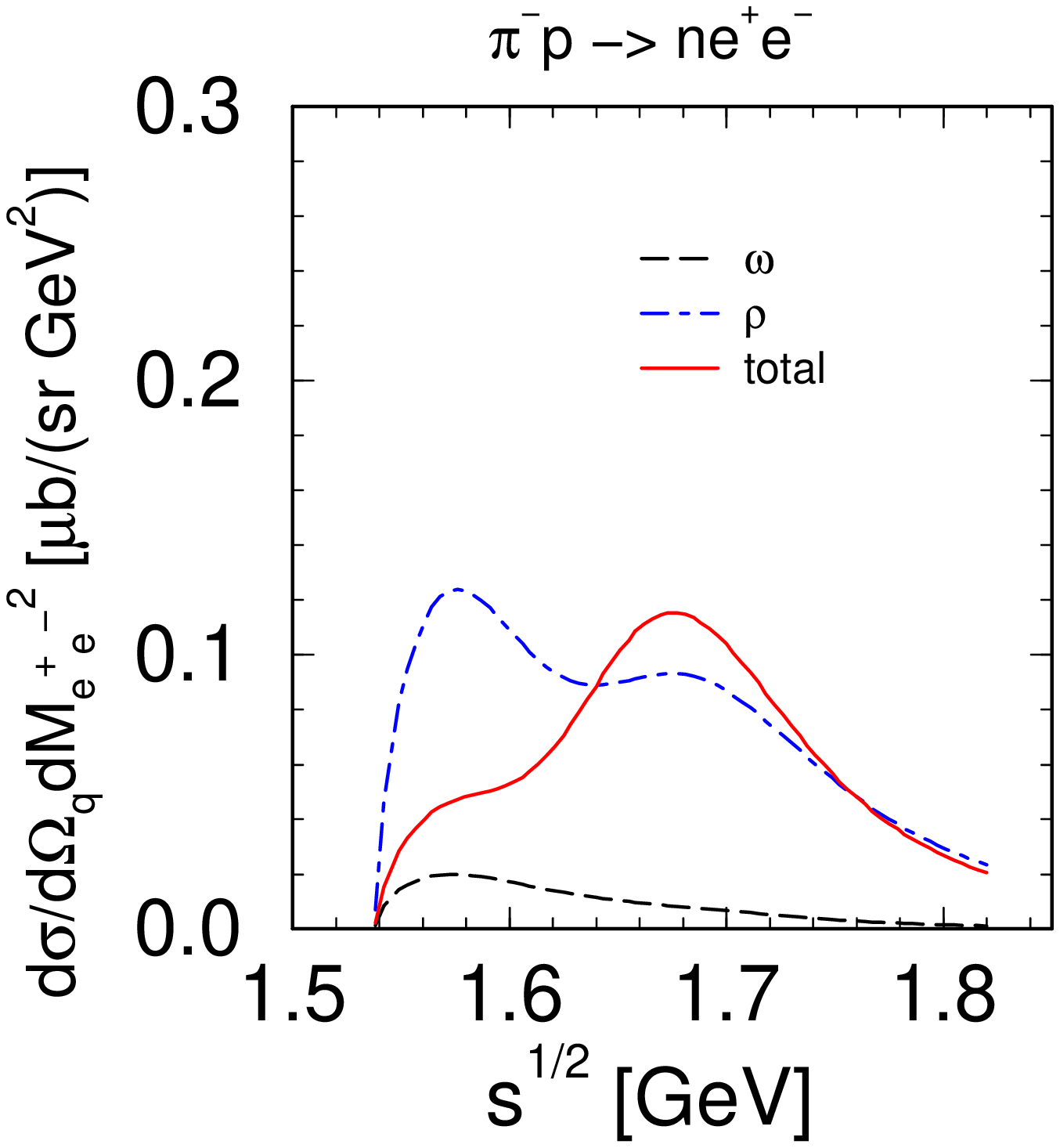, width=7.cm}\qquad\qquad
 \epsfig{file=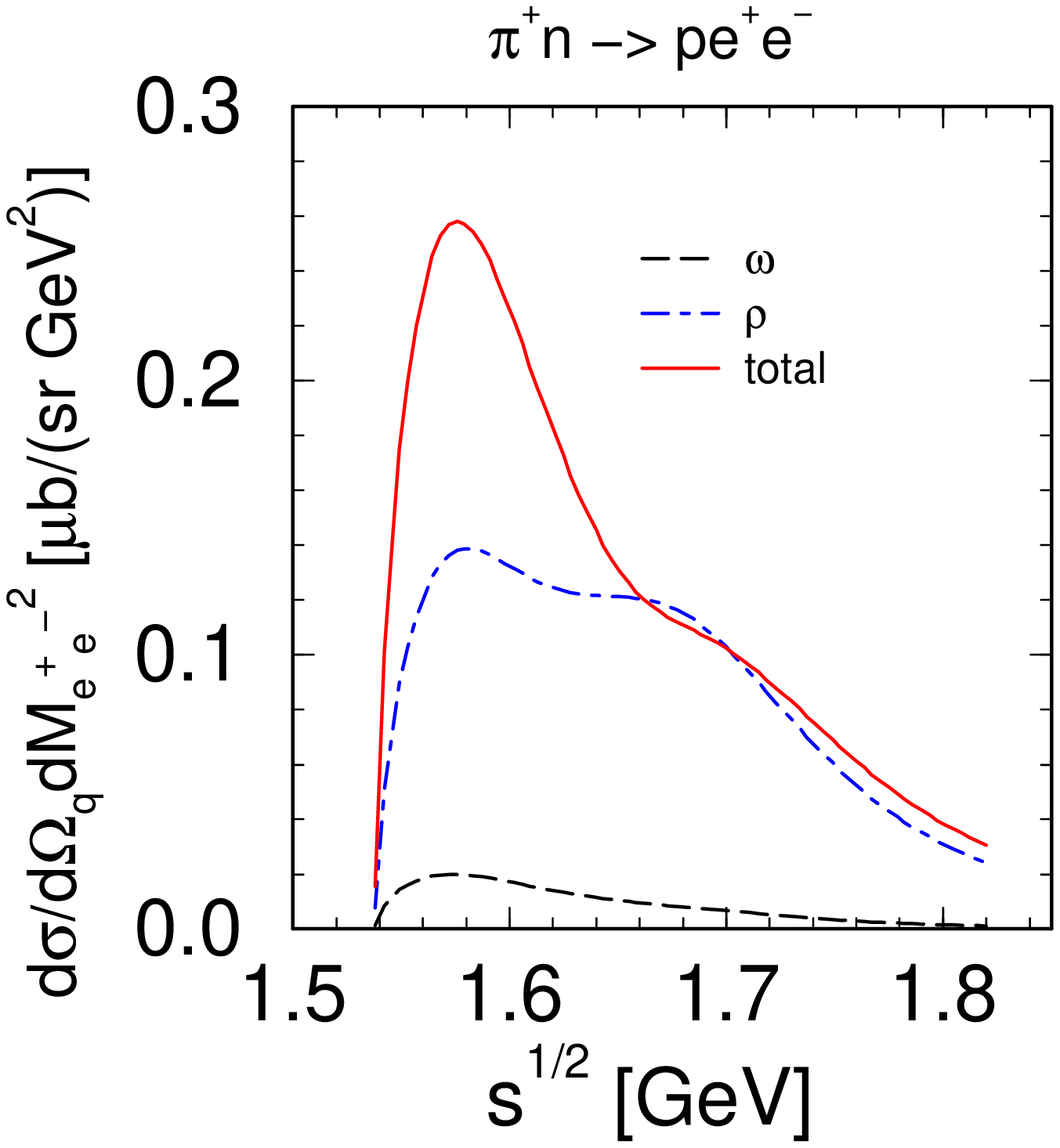, width=7.cm}}
\caption{
Differential cross sections of dielectron production
as a function of $s^{1/2}$ for $M_{e^+e^-}=0.6$ GeV
for the reactions $\pi^- p\to ne^+e^-$ (left panel) and
$\pi^+ n\to pe^+e^-$ (right panel).}
\label{fig:7}
\end{figure}

\begin{figure}
\centering
{\epsfig{file=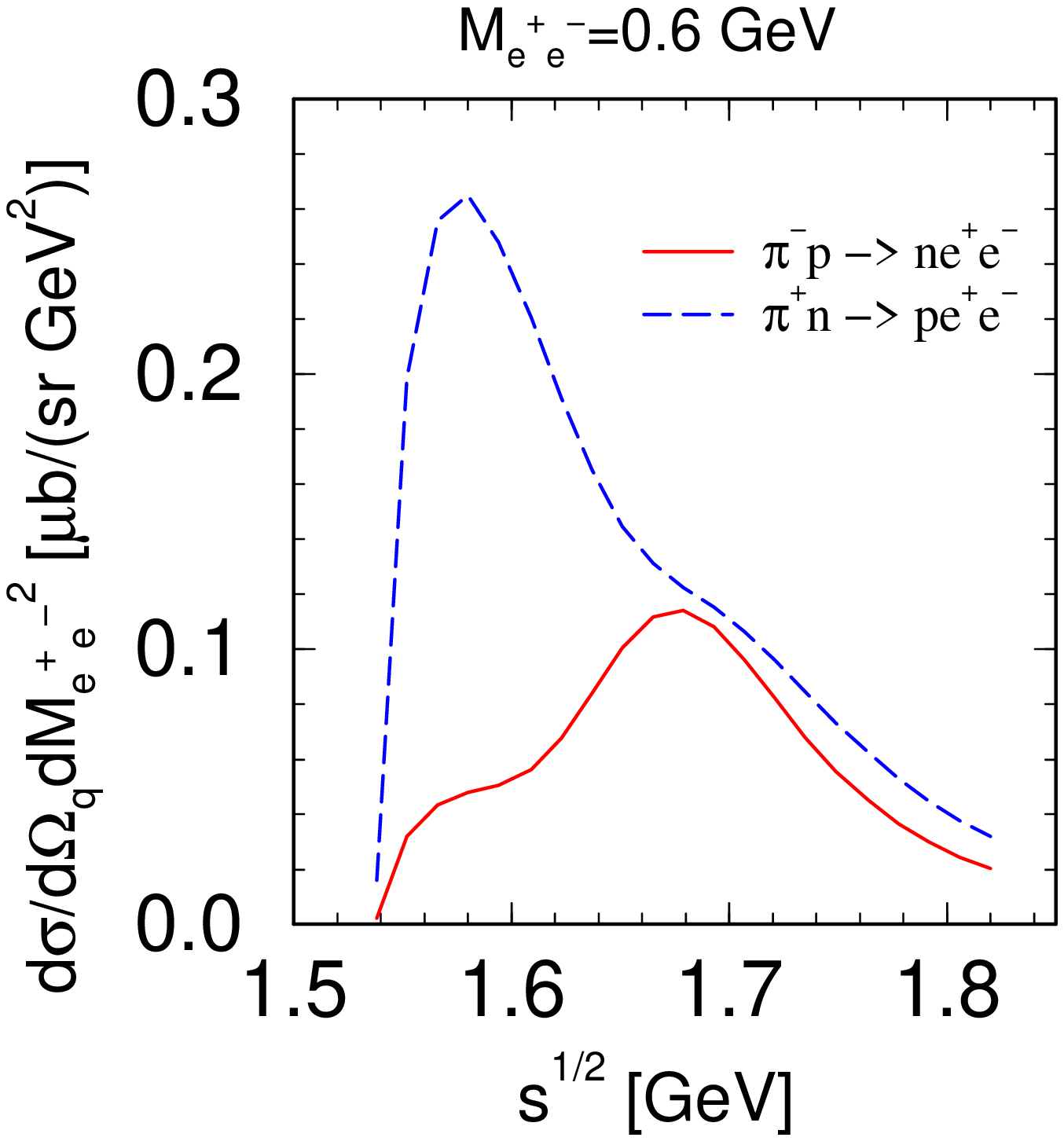, width=7.cm}\qquad\qquad
 \epsfig{file=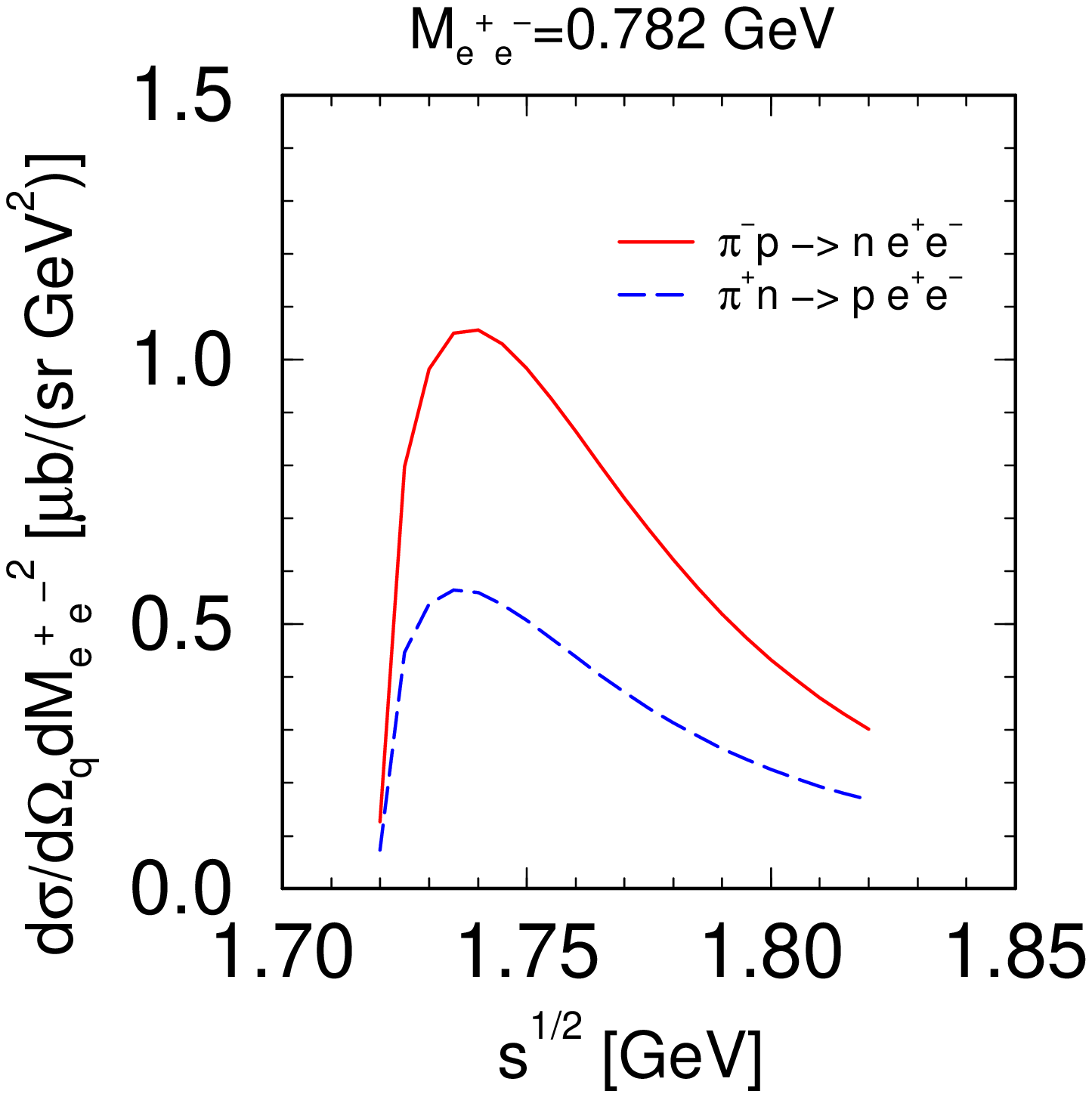, width=7.45cm}}
\caption{
Differential cross section of dielectron production
for the reactions
$\pi^- p \to n e^+ e^-$  and $\pi^+ n \to p e^+ e^-$ as a function
of $s^{1/2}$ for $M_{e^+e^-}=0.6$ GeV (left panel)
and 0.782  GeV (right panel).}
\label{fig:8}
\end{figure}

\begin{figure}
\centering
{\epsfig{file=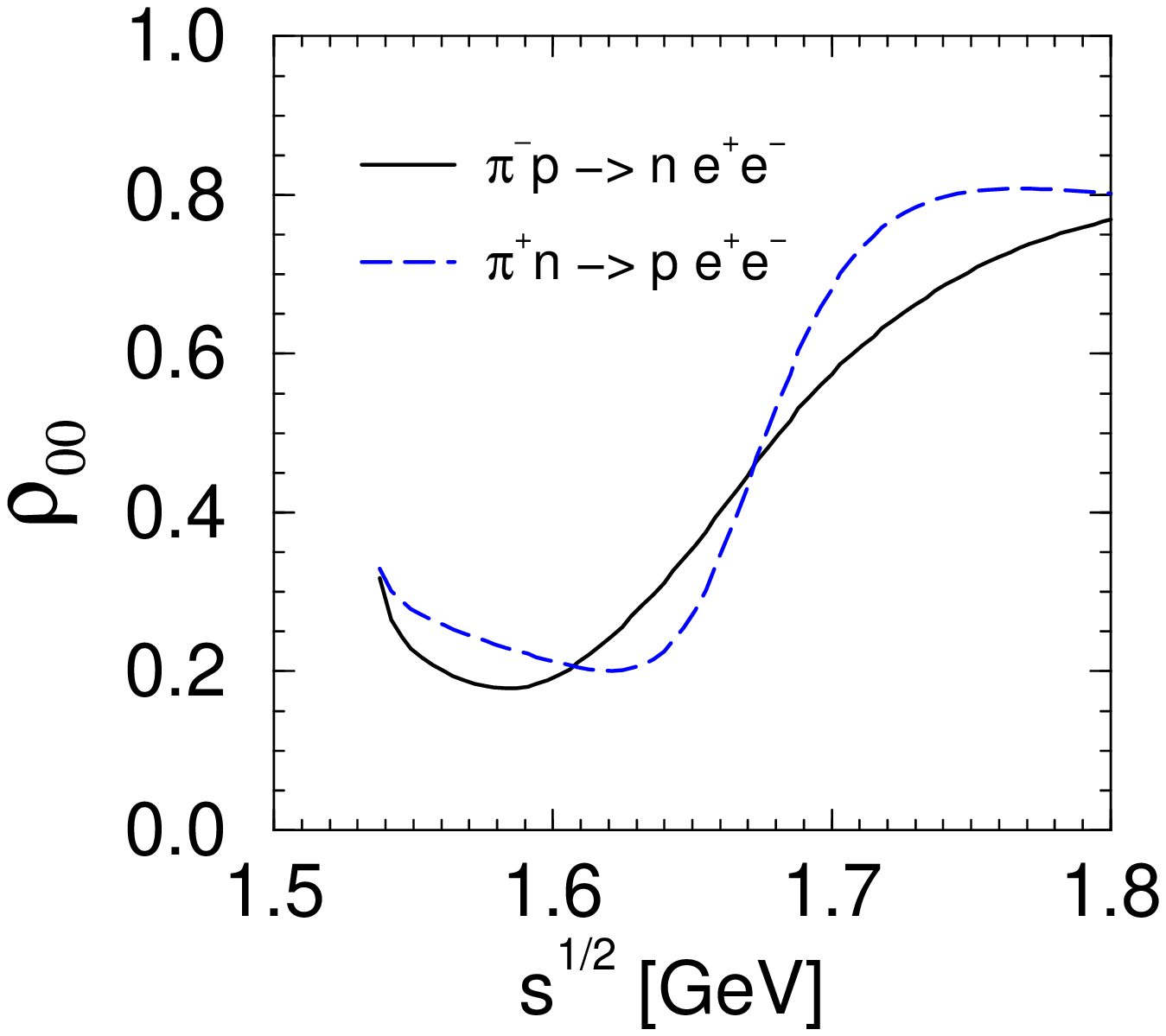, width=7.cm}\qquad\qquad
 \epsfig{file=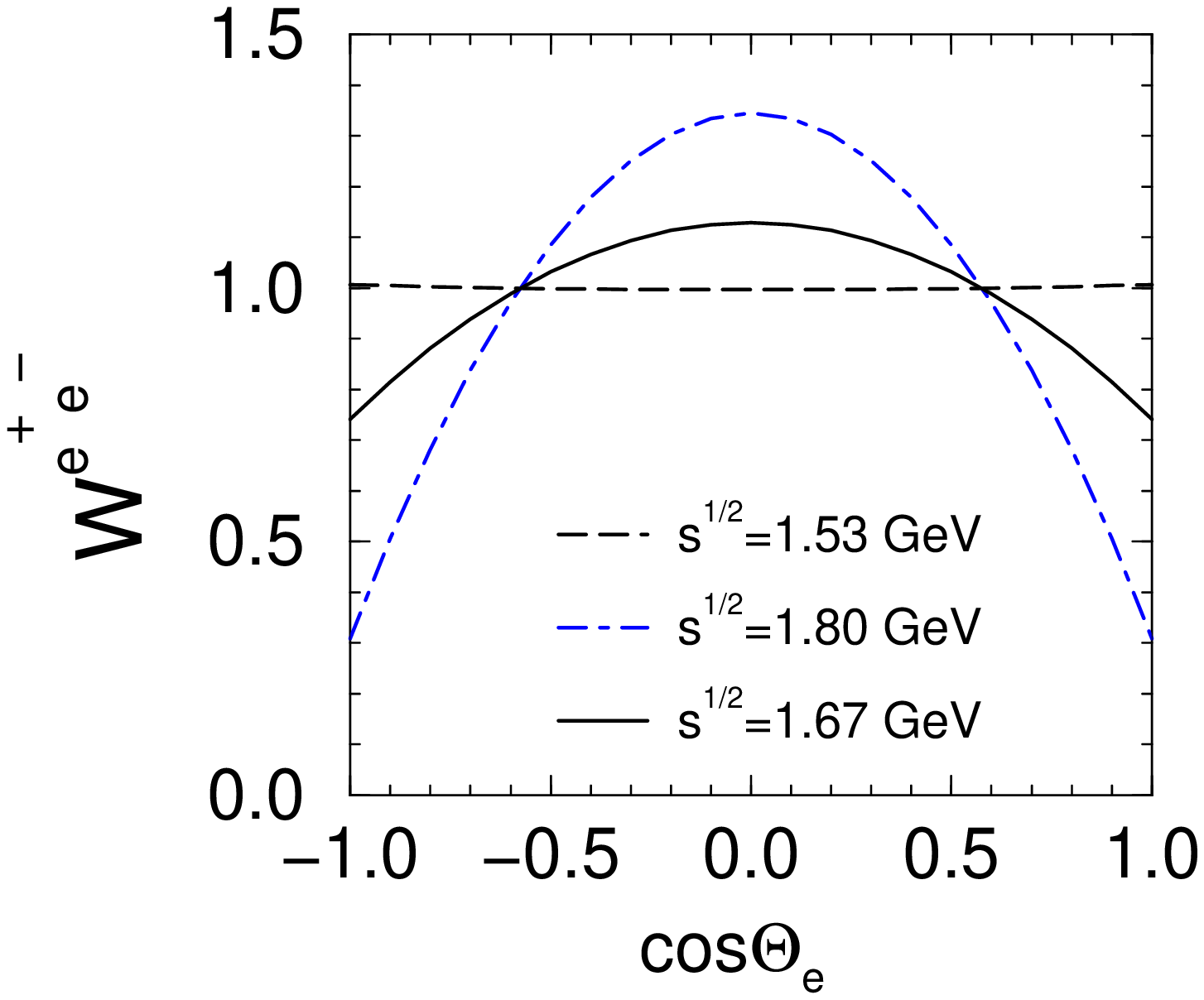, width=7.cm}}
\caption{
Spin density matrix element $\rho_{00}$ for
$\pi^- p\to n e^+e^-$ (solid lines) and $\pi^+ n\to p e^+e^-$
(dashed lines) as a function of
$s^{1/2}$  for $M_{e^+e^-}=0.6$ GeV (left panel), and the
angular distributions of electrons at $M_{e^+e^-}=0.6 $
and  different values of $s^{1/2}$ (right panel). 
}
\label{fig:9}
\end{figure}

\begin{figure}
\centering
{\epsfig{file=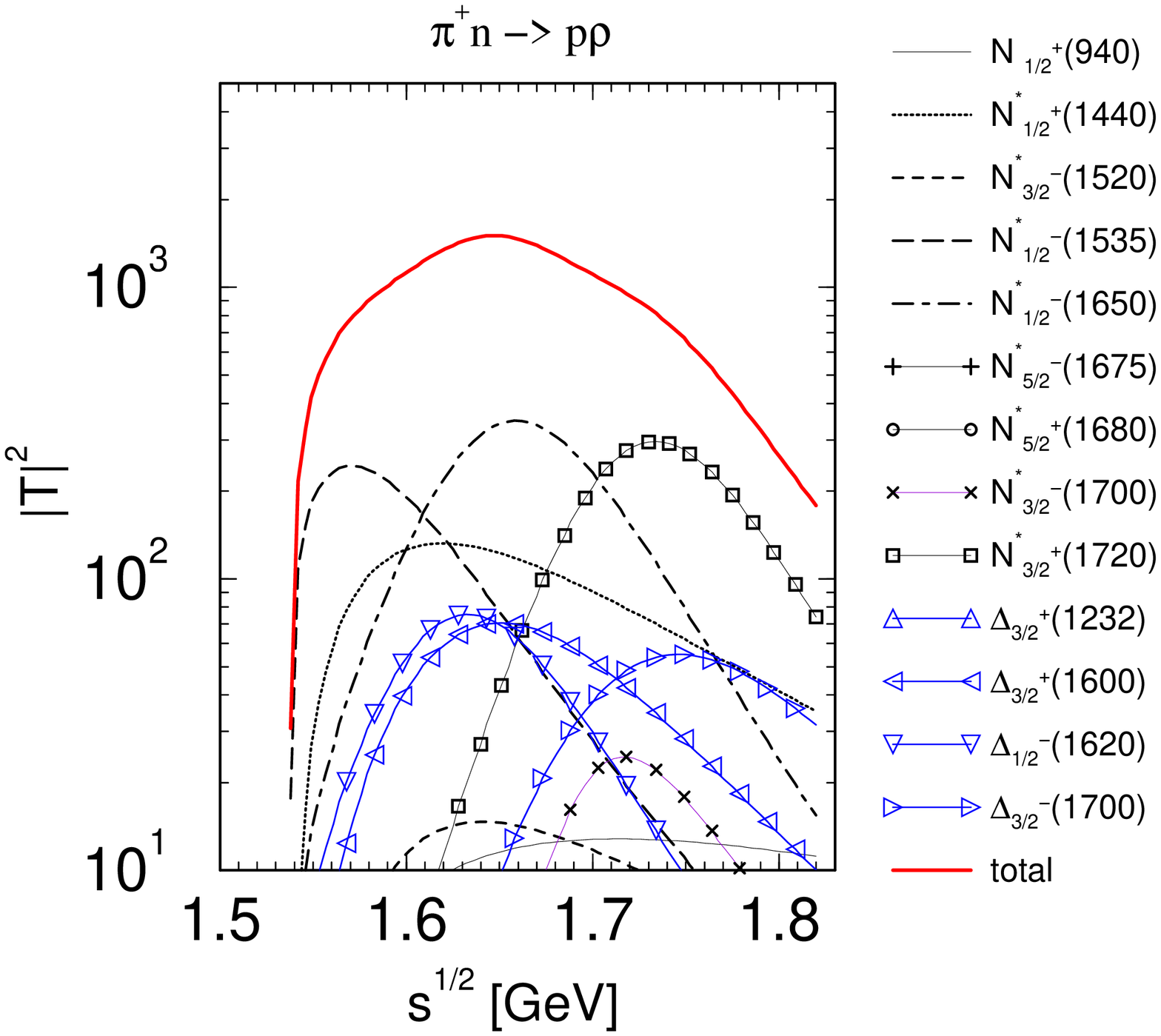, width=7.2cm}\qquad\qquad
 \epsfig{file=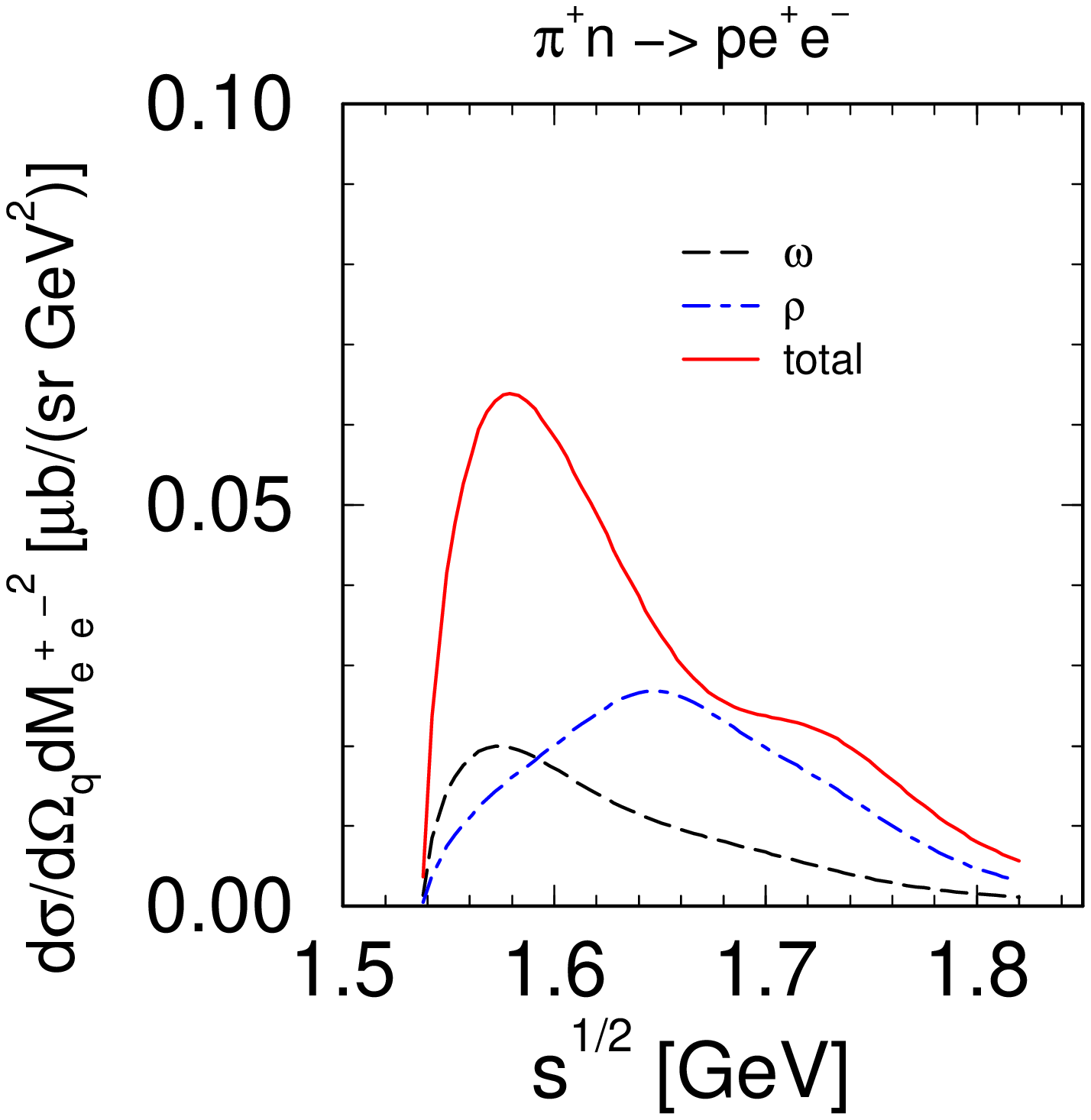 , width=6.8cm}}
\caption{
Left panel:
individual contributions of nucleon resonances listed in Table~2
to the spin averaged invariant amplitude
of $\rho$ production
at  $M_{e^+e^-} = 0.6$ GeV.
Right panel:
the same as in the right panel of Fig.~7
but with resonance parameters listed in Table~2.}
\label{fig:10}
\end{figure}

\begin{figure}
\centering
{\epsfig{file=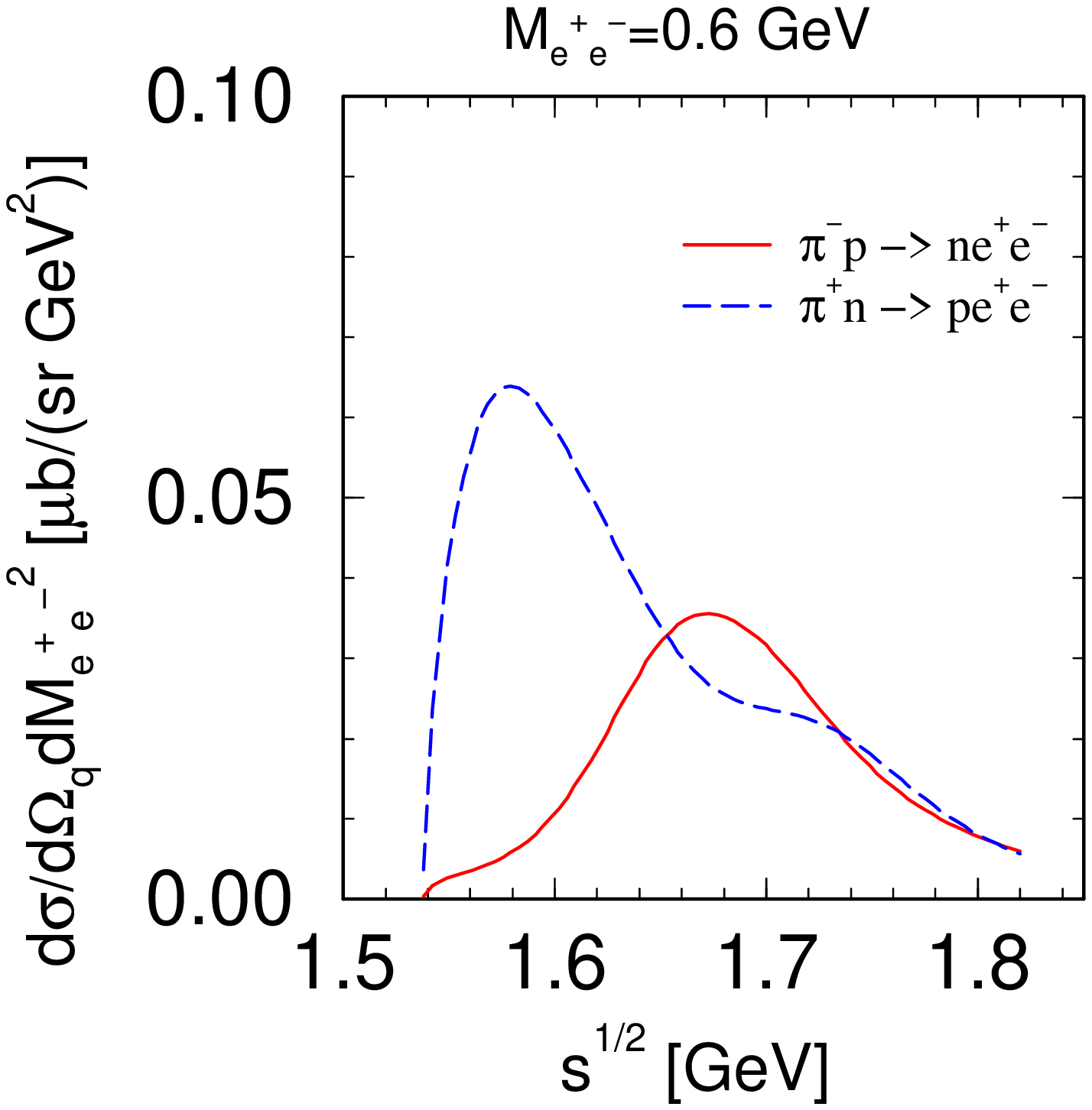, width=7.cm}\qquad\qquad
 \epsfig{file=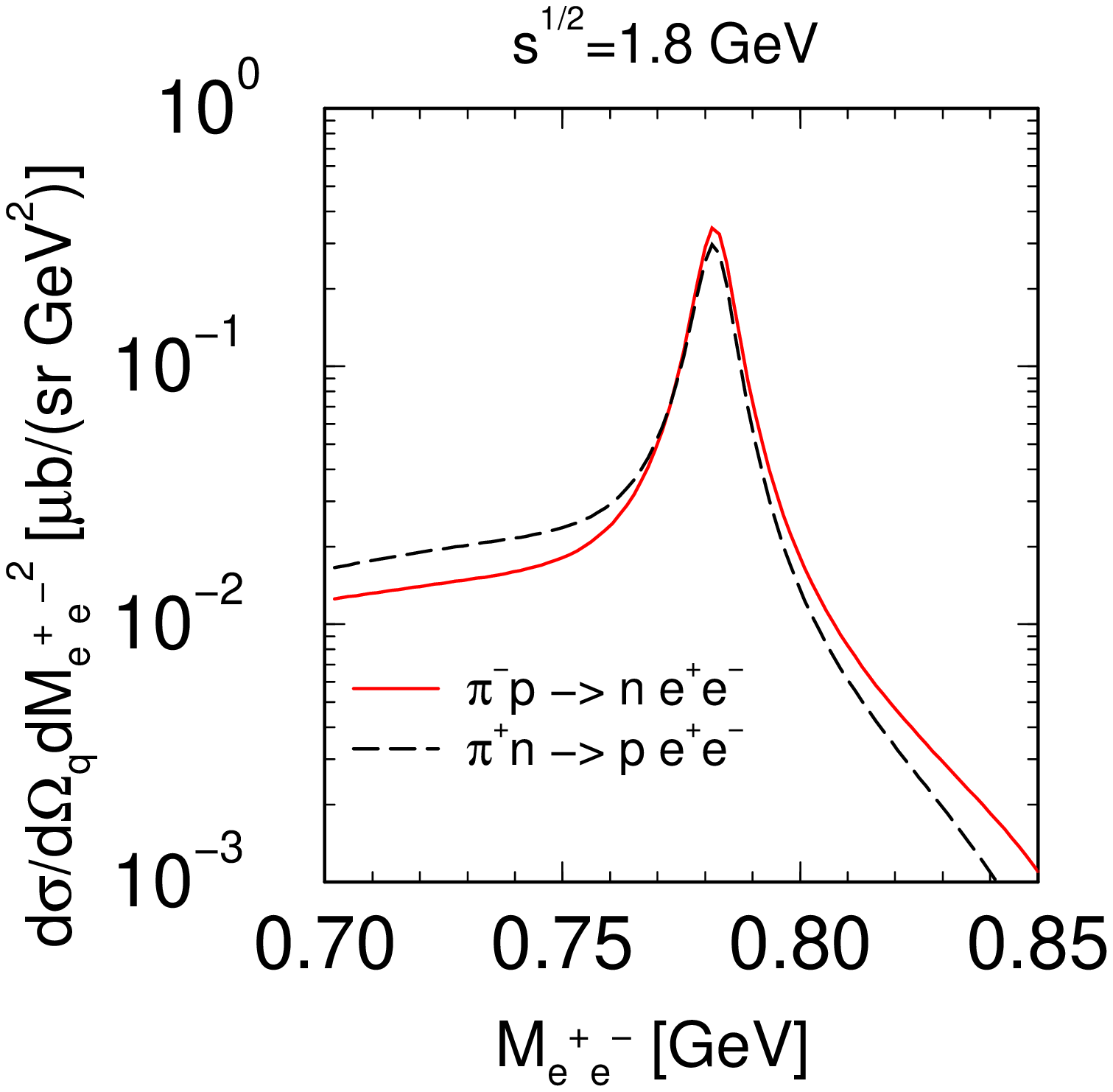, width=7.3cm}}
\caption{
Left panel:
the same as in the left panel of Fig.~8 but
but with resonance parameters listed in Table~2.
Right panel: the same as in the right panel in Fig.~3
but with resonance parameters listed in Table~2.}
\label{fig:11}
\end{figure}

\begin{figure}
\centering
{\epsfig{file=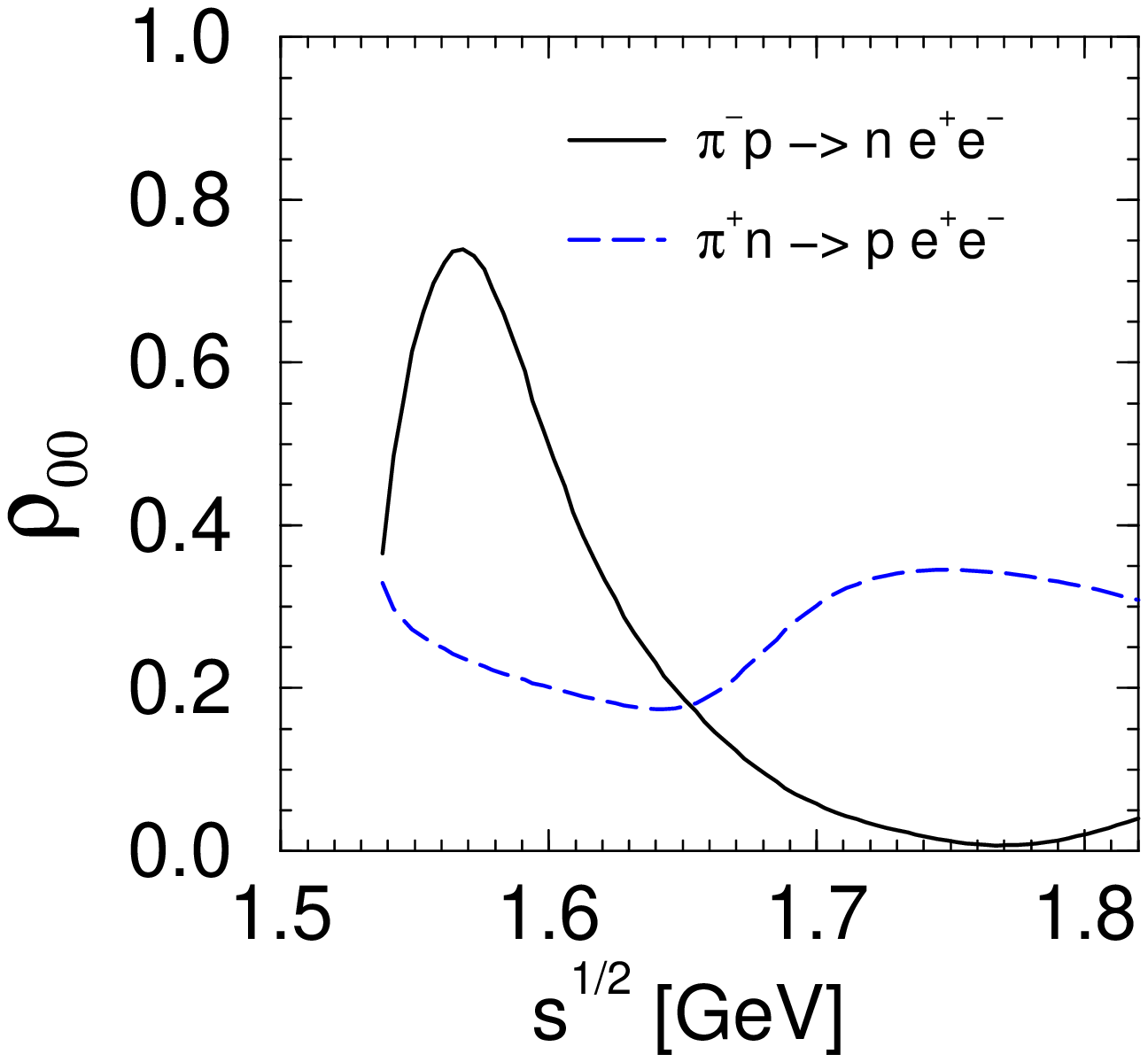,width=6.7cm}\qquad\qquad
 \epsfig{file=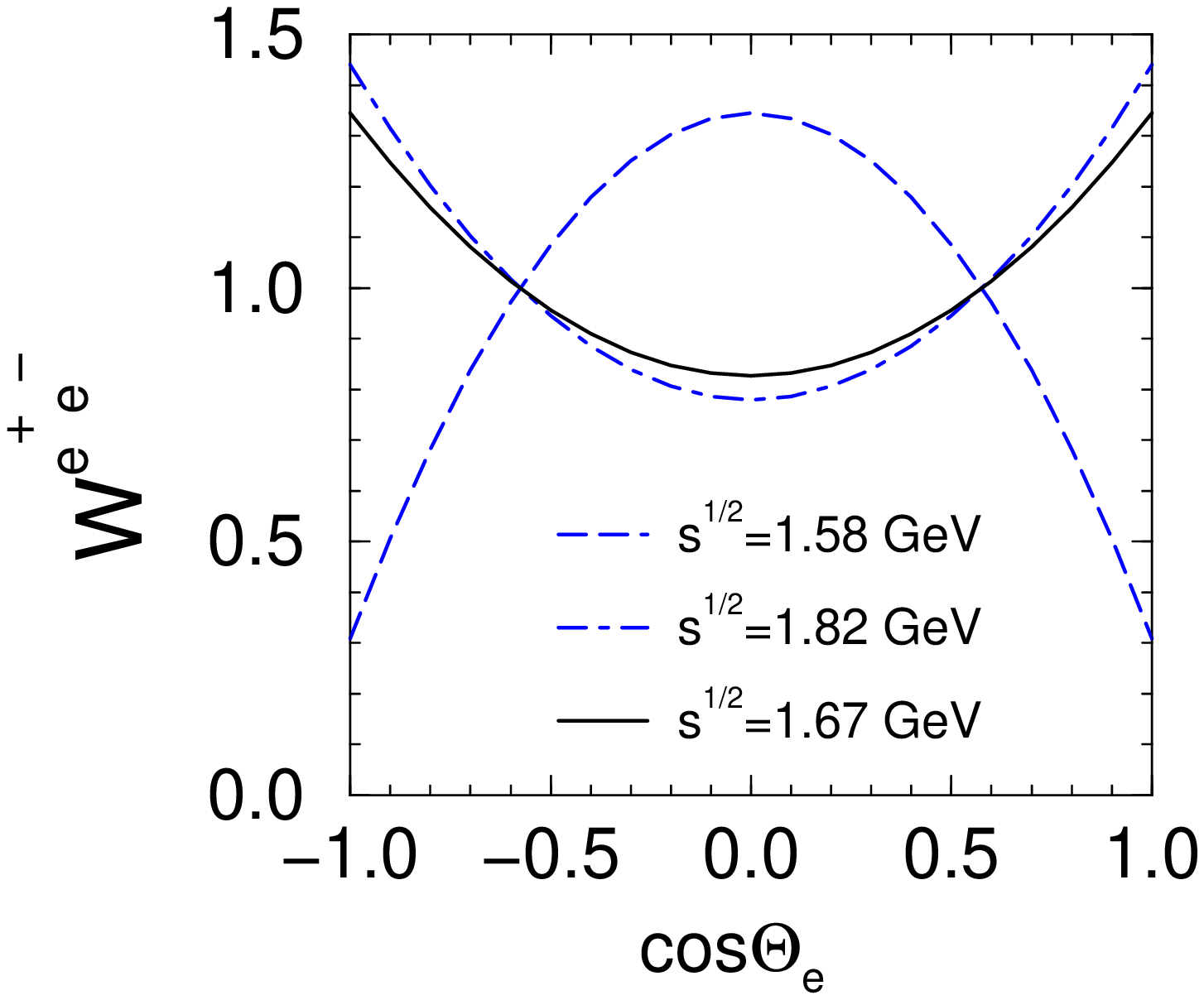,width=7.cm}}
\caption{
The same as in Fig.~9
but with resonance parameters listed in Table~2.}
\label{fig:12}
\end{figure}

\end{document}